\documentclass[conference]{IEEEtran}
\IEEEoverridecommandlockouts
\usepackage{ifpdf}
\usepackage[latin1]{inputenc}
\usepackage{amsfonts}
\usepackage{amsmath}
\usepackage{mathrsfs}
\usepackage{balance}
\usepackage{cite}
\usepackage{subfigure}

\usepackage{listings}
\usepackage{color}
\usepackage{units}
\usepackage{nicefrac}
\usepackage[printonlyused]{acronym}
\usepackage{pstricks, pst-plot, pst-grad, pstricks-add, pst-node, pstricks-add}
\usepackage[section] {placeins}

\usepackage{ifpdf}
\ifCLASSINFOpdf
   \usepackage[pdftex]{graphicx}
   \graphicspath{{./figures/pdf/}}
   \DeclareGraphicsExtensions{{.pdf}}
\else
   \usepackage[dvips]{graphicx}
   \graphicspath{{./figures/eps/}}
   \DeclareGraphicsExtensions{{.eps}}
\fi

\usepackage{hyperref}

\acrodef{FEC}[FEC]{forward error correction}
\acrodef{MCS}[MCS]{modulation and coding scheme}
\acrodef{PDF}[pdf]{probability density function}
\acrodef{cdf}[cdf]{cumulative density function}
\acrodef{RAN}[RAN]{radio access network}
\acrodef{RRH}[RRH]{remote radio head}
\acrodef{SNR}[SNR]{signal-to-noise ratio}
\acrodef{SINR}[SINR]{signal-to-interference-and-noise ratio}
\acrodef{C-RAN}[C-RAN]{centralized radio access network} 
\acrodef{CB}[CB]{code block}
\acrodef{TB}[TB]{transport block}
\acrodef{CBLER}[CBLER]{code block error rate}
\acrodef{TBLER}[TBLER]{transport block error rate}
\acrodef{VM}[VM]{virtual machine}
\acrodef{RAP}[RAP]{radio access point}
\acrodef{RB}[RB]{resource block}
\acrodef{FLOPS}[FLOPS]{floating-point operations per second}
\acrodef{PPP}[PPP]{Poisson point process}
\acrodef{CP}[CP]{cloud processing}
\acrodef{LP}[LP]{local processing}
\acrodef{MRS}[MRS]{max-rate selection}
\acrodef{CAS}[CAS]{computationally aware selection}
\acrodef{BS}[BS]{base station}

\newrgbcolor{darkgreen}{0.2 0.6 0.2}
\newrgbcolor{darkred}{0.6 0.2 0.2}
\newrgbcolor{darkblue}{0.2 0.2 0.6}

\begin{document}
\title{The Role of Computational Outage in Dense \\  Cloud-Based Centralized Radio Access Networks}
\author{\IEEEauthorblockN{Matthew C. Valenti\IEEEauthorrefmark{1}, Salvatore Talarico\IEEEauthorrefmark{1}, and Peter Rost\IEEEauthorrefmark{2}}
\IEEEauthorblockA{\IEEEauthorrefmark{1}West Virginia University, Morgantown, WV, USA. \\
\IEEEauthorrefmark{2}NEC Laboratories Europe, Heidelberg, Germany}
\thanks{The research leading to these results has received partly funding from the European Union Seventh Framework Programme (FP7/2007-2013) under grant agreement n\textordmasculine~317941 (www.ict-ijoin.eu).
  The authors would like to acknowledge the contributions of their colleagues in iJOIN, although the views expressed are those of the authors and do not necessarily represent the project.}
}
%
%

\date{}
\maketitle

\begin{abstract}
Centralized radio access network architectures consolidate the baseband operation towards a cloud-based platform, thereby allowing for efficient utilization of computing assets, effective inter-cell coordination, and exploitation of global channel state information.   This paper considers the interplay between computational efficiency and data throughput that is fundamental to centralized RAN.  It introduces the concept of computational outage in mobile networks, and applies it to the analysis of complexity constrained dense centralized RAN networks.  The framework is applied to single-cell and multi-cell scenarios using parameters drawn from the LTE standard.  It is found that in computationally limited networks, the effective throughput can be improved by  using a computationally aware policy for selecting the modulation and coding scheme, which sacrifices spectral efficiency in order to reduce the computational outage probability.  When signals of multiple base stations are processed centrally, a computational diversity benefit emerges, and the benefit grows with increasing user density.
\end{abstract}
\IEEEpeerreviewmaketitle
%
%
\section{Introduction}
  In an information society as we have it today, mobile voice and data communication is a commodity service that is supposed to be available everywhere at any time.
  Accordingly, novel technologies to improve system capacity and quality of service are proposed and discussed in the context of next generation mobile networks.
  However, the focus of this discussion has to date focused primarily on the system performance, while the increased demand for computational resources has received relatively little attention.

  \subsection{Computational requirements in mobile networks}
    Mobile networks are hard real-time systems with tight timing and protocol constraints. These constraints are described by mobile communications standards such
    as 3GPP LTE \cite{3GPP.website} and must not be violated. In order to fulfill these constraints, a pre-defined amount of computational resources
    must be provided such that downlink and uplink processing can be performed within a given time interval. For instance, the uplink \ac{MCS} determines
    the number of information bits that must be processed per subframe.

    The current trend is towards the densification of mobile networks, where smaller cells are deployed. Compared to traditional macro-cellular base stations,
    small-cell base stations will be characterized by fewer antennas, different antenna and wave propagation patterns, lower transmission power, and limited computational
    resources. The latter characteristic is a consequence of economic constraints and the need for compact radio access points.
    Limiting the amount of computational resources may lead to \emph{computational} outage rather than channel outage.  Such computational outages occur whenever a decoding failure occurs due to the violation of a timing constraint rather than due to insufficient channel conditions.

    Computational outage leads to a waste of spectral resources and loss of throughput, in much the same way as a channel outage.  It is feasible that for some deployments, computational outage could be the dominant form of outage, and as such, will influence the design of the scheduler and certain aspects of the network architecture.   The use of a \emph{computationally aware} scheduler could yield great benefits when used in a network with computational constraints.  Furthermore, the pooling of computational assets in a \emph{cloud} could yield a \emph{computational diversity} benefit, whereby centralized processing elastically focuses resources on those cells that have high momentary computing requirements.

  \subsection{Related work}
    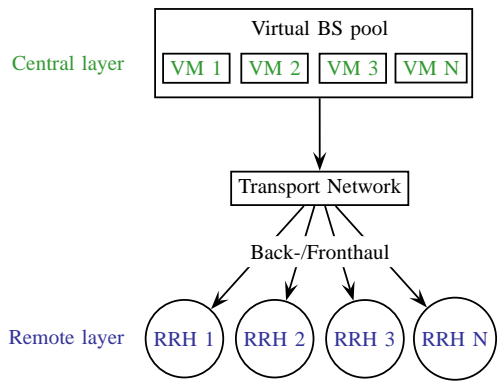
\begin{figure}
      \centering
      \scalebox{0.8}{\begingroup
\unitlength=1mm
\begin{picture}(79, 62)(0, 0)
  \psset{xunit=1mm, yunit=1mm, linewidth=0.3mm}
  \psset{arrowsize=3pt 4, arrowlength=1.4, arrowinset=.4}
  \rput(0, -3){%
    \rput(17, 0){%
      \cnodeput(10, 10){RRH1}{\textcolor{darkblue}{RRH 1}}%
      \cnodeput(25, 10){RRH2}{\textcolor{darkblue}{RRH 2}}%
      \cnodeput(40, 10){RRH3}{\textcolor{darkblue}{RRH 3}}%
      \cnodeput(55, 10){RRH4}{\textcolor{darkblue}{RRH N}}%
      \rput[r](0, 10){\textcolor{darkblue}{Remote layer}}%
      \rput[c](32.5, 35){\rnode{Transport}{\psframebox[linecolor=black]{Transport Network}}}%
      \rput(12, 55){\psframebox[linecolor=black]{\textcolor{darkgreen}{VM 1}}}%
      \rput(25, 55){\psframebox[linecolor=black]{\textcolor{darkgreen}{VM 2}}}%
      \rput(38, 55){\psframebox[linecolor=black]{\textcolor{darkgreen}{VM 3}}}%
      \rput(51, 55){\psframebox[linecolor=black]{\textcolor{darkgreen}{VM N}}}%
      \pnode(32.5, 50){VM}\psframe(5, 50)(58, 65)%
      \rput[t](32.5, 63){Virtual BS pool}%
      \rput[r](0, 55.5){\textcolor{darkgreen}{Central layer}}%
      \ncline{->}{VM}{Transport}%
      \ncline{->}{Transport}{RRH1}%
      \ncline{->}{Transport}{RRH2}%
      \ncline{->}{Transport}{RRH3}%
      \ncline{->}{Transport}{RRH4}%
      \rput(32.5, 24.5){\psframebox[linecolor=white, fillcolor=white, fillstyle=solid]{Back-/Fronthaul}}%
      }%
  }
\end{picture}
\endgroup
      }
      \caption{Typical cloud RAN architecture}
      \label{fig:system.model:cloud.architecture}
    \end{figure}
    The deployment of very dense networks requires novel technologies to allow for improved interference coordination and to improve the utilization of networks.
    Already today, networks are underutilized at less than \unit[40]{\%} maximum load. One technology to improve interference coordination and network utilization
    is centralized \ac{RAN}
    \cite{Wang.ChinaComm.2010, Guan.Kolding.Merz.CRAN.2010,Rost.etal.ComMag.2014}. Centralized \ac{RAN} achieves this improvement by centralizing all baseband processing to a central
    entity. Currently deployed centralized \ac{RAN} solutions exploit utilization improvements on a base station
    level; i.\,e., each baseband unit in a  pool of units may be statically assigned to a particular \ac{RRH}.

    Using cloud computing as a platform for centralized \ac{RAN} allows for the use of
    virtualization technologies that offer flexible provisioning of computational resources. Consider Fig. \ref{fig:system.model:cloud.architecture},
    where the \ac{RAN} is divided into a remote and central layer. Depending on the backhaul quality, parts of the radio protocol stack are centralized and
    executed within a virtual \ac{BS} pool \cite{Rost.etal.ComMag.2014}. Each \ac{VM} within this virtual \ac{BS} pool may represent a \ac{RAP} or cluster of
    \acp{RAP}, where each \ac{RAP} may be a fully centralized base station (\ac{RRH}) or just a partially centralized base station.
    Based on the actual traffic and resulting computational demand, each \ac{VM} is provisioned with resources that can be elastically adjusted.
    One such split between remote and central layer could be to perform operations common to all users of a BS (e.g. FFT) at the \ac{RAP} and to perform user-specific
    operations such as \ac{FEC} at the central layer  \cite{Doetsch.Doll.Mayer.etal.BellLabs.2013, Rost.etal.ComMag.2014}. \ac{FEC} consumes a major part of the computational resources,
    particularly in the uplink \cite{Bhaumik.Chandrabose.Jataprolu.Kumar.Muralidhar.Polakos.Srinivasan.Woo.MobiCom.2012}.
    Hence, centralizing \ac{FEC} (and all functionality above) allows for centralizing a large part of the base station complexity and provides opportunities to exploit
    the multi-user computational diversity.

    However, in order to exploit such a cloud-computing platform efficiently, new technologies are needed to  predict, monitor, and control the computational
    requirements. This requires the joint operation and optimization of the mobile network's communication and computation systems.
    Recently, the computational complexity of mobile communication has received increasing attention, in particular for short-distance communication where power consumption due to
    transmitter and receiver processing may be on the order of the transmission power.
    In \cite{Grover.Woyach.Sahai.JSAC.2011}, Grover~\textit{et al.} showed that decoding complexity would scale at moderate to high \ac{SNR} with $O(\log(x)^{-1})$ in
    the ratio of necessary \ac{SNR} to achieve capacity and necessary \ac{SNR} to achieve the chosen data rate.
    Hence, to get closer to capacity, the required complexity increases super-linearly.

  \subsection{Contribution and Outline}
    When small cells are provided with limited computational resources or multiple small cells share the (virtualized) computational resources, computational
    outage becomes increasingly important. However, determining computational outage is very difficult as it requires a detailed complexity model (similar to channel outage),
    and computational outage depends on multiple parameters such as \ac{SNR}, block-length, modulation-and-coding-scheme, and decoder implementation.

    In this paper, we analyze the likelihood of computational outage in a mobile communication network. More specifically, we first derive a model for computational outage
    and show how it relates to channel outage.
    In order to evaluate the actual impact on the throughput performance,
    this model is applied to a block Rayleigh fading channel and a network with co-channel interference.  Local processing of the uplink signal is compared with central processing, and the computational diversity benefit of central processing is noted.
    We further introduce
    a computationally aware \ac{MCS} selection policy that reduces the computational complexity requirements at the cost of slightly decreased spectral efficiency. However, in complexity constrained deployments, the selection policy provides higher  effective throughput.

    The paper is structured as follows: Section \ref{sec:channel.computational.outage} introduces the notion of computational outage and relates it to channel outage,
    Section \ref{sec:outage.single.cell} analyzes the outage behavior in the case of a single cell, Section \ref{sec:outage.network}
    extends the analysis to a multi-cell environment where co-channel interference has to be considered. The paper is concluded in Section \ref{sec:conclusions}.

\section{Channel and Computational Outage}\label{sec:channel.computational.outage}
Consider a {\em cloud group} containing $N_\mathsf{cloud}$ base stations or \emph{\acp{RAP}} whose signals are jointly processed in a virtual BS pool.  Let $Y_i$ indicate the $i^{th}$ RAP and its location within the network.  When a mobile in the cell served by $Y_i$ transmits, a \ac{TB} is received at $Y_i$ with a \ac{SINR} equal to $\gamma_i$.  The \ac{SINR} is assumed to be fixed for the duration of the subframe, which is true if the desired and all the interfering signals are subject to block fading\footnote{In the present paper, we use block fading as it eases the exposition. Fast correlated fading is for future study.}, and hence the channel is AWGN with \ac{SNR} $\gamma_i$.

When a \ac{TB} is received with \ac{SINR} $\gamma$, there is some probability that it will not be correctly decoded.   Ideally, it will always be correctly decoded if $\gamma$ is above some threshold and will be incorrectly decoded if it is below that threshold (a ``brick wall'' error-rate curve).  However, this behavior requires infinitely long codes.  In practice, a finite-length code, such as a turbo code, must be used, and its error-rate curve will not instantaneously drop to zero.  To model this effect, we define a {\em random} error-indicator function $E(\gamma)$ which returns a 1 if a particular TB received with \ac{SINR} $\gamma$ fails to be successfully decoded and a 0 if it is correctly decoded.  A {\em channel} outage occurs whenever $E(\gamma)=1$.  It follows that the channel outage probability is
\begin{eqnarray}
\epsilon_\mathsf{channel} & = & \mathbb P [E(\gamma)=1] = \mathbb E [  E(\gamma) ].
\end{eqnarray}

As the TB format and the number of turbo decoder iterations will vary from one transmission to the next, we furthermore define a random function $\mathcal C(\gamma)$ to be the \emph{computational effort} required to process a particular TB received with \ac{SINR} $\gamma$.   Unless the link layer decides to drop a TB to prevent a computational overflow, every received TB must be processed.  Thus, even if a channel outage occurs, $\mathcal C(\gamma)$ will generally be nonzero.  However, using an early-stopping criteria in the decoder may allow $\mathcal C(\gamma)$ to be less than its maximum value when a channel outage occurs.



The virtual BS pool will be provisioned with a limited amount of computational resources $\mathcal C_\mathsf{max}$ per RAP. The units of these resources are the same as those associated with the computational effort $\mathcal C(\gamma)$.  If the total required computational effort exceeds $N_\mathsf{cloud} \cdot \mathcal C_\mathsf{max}$ before a decoding deadline is reached, a \emph{computational} outage will occur.  It follows that the computational outage probability for the cloud group is
\begin{eqnarray}
   \epsilon_\mathsf{comp}
   & = &
   \mathbb P
   \left[
       \sum_{i=1}^{N_\mathsf{cloud}} \mathcal C(\gamma_i) > N_\mathsf{cloud} \cdot \mathcal C_\mathsf{max}
   \right].  \label{Eqn:Complexity}
\end{eqnarray}
When a computational outage occurs, not all of the TBs in the cloud group are necessarily lost.  If the processing is suitably scheduled, it is possible for some of the TBs to be correctly decoded, though there will not be enough resources to correctly decode all of the TBs in the group. The actual set of TBs that are lost depends highly on the deployed task scheduling algorithm.  Thus, a TB can be lost due to a channel outage or a computational outage.  Since even those TBs that are in a channel outage must be processed, these two kinds of outages are not mutually exclusive.  Indeed, it is possible for a TB in a channel outage to trigger a computational outage as the baseband processor attempts to decode it.   The \emph{overall} outage probability $\epsilon$, or simply \emph{outage} probability, is the probability that a TB is lost due to either kind of outage.

Due to fluctuations in channel quality and traffic load, the SINRs at the different RAPs in the cloud group will generally be quite different, and as a consequence the offered computational loads $\mathcal C(\gamma_i), i \in \{1, ..., N_\mathsf{cloud}\},$ will vary from RAP to RAP.  Centralizing the processing exploits this diversity of computational load, as computational assets can be diverted from RAPs with a currently low computational load to those with a high computational load.  This \emph{computational diversity} is a key benefit of cloud-based centralized radio access networks.

\section{Outage in an Isolated Cell}\label{sec:outage.single.cell}
Consider a network with just one cell containing a single active user served by a single RAP.   There is no interference from other cells, and the signal received by the RAP is processed all by itself; hence, $N_\mathsf{cloud}=1$.  The statistics of $E(\gamma)$ and $\mathcal C(\gamma)$ are found through simulation.   To provide meaningful results, the simulations used to determine these statistics are executed using parameters from the LTE standard.

\subsection{Complexity-Throughput Tradeoffs in LTE}

LTE features adaptive modulation and coding using turbo codes and hybrid ARQ.  The RAP (called an \emph{eNodeB}) commands the mobile (called a \emph{UE}) to transmit using one of 27 distinct \acp{MCS}.  Each MCS is identified by an MCS index, $I_\mathsf{mcs} = \{0,...,26\}$, and is characterized by a different combination of code rate and modulation format \cite{3GPP.TS.36.211, 3GPP.TS.36.212}.  Three kinds of modulation are used: QPSK ($0\leq I_\mathsf{mcs} \leq 10$), 16-QAM ($11\leq I_\mathsf{mcs} \leq 20$), and 64-QAM ($21\leq I_\mathsf{mcs} \leq 26$).  When a \ac{TB} is larger than 6144 bits, it is segmented into multiple \emph{\acp{CB}}.  Each CB is separately turbo encoded, and all $C$ CBs in the TB must be correctly decoded for the TB to be correct.    Let $\epsilon_\mathsf{cb}$ be the probability that a CB is in a channel outage.  It follows that $\epsilon_\mathsf{channel} = 1-(1-\epsilon_\mathsf{cb})^C$ is the probability that the TB is in a channel outage.



Turbo decoding is a computationally demanding task.  However, because decoding is iterative and stops when a CRC check is satisfied, the amount of computational effort devoted to turbo decoding is highly variable. To handle this challenge, a centralized \ac{RAN} deployment could virtualize its processing resources and flexibly assign turbo decoding tasks to
available computing resources.
Ultimately, it is the turbo decoding that dominates the computational outage behavior because it consumes a major part of baseband processing resources \cite{Bhaumik.Chandrabose.Jataprolu.Kumar.Muralidhar.Polakos.Srinivasan.Woo.MobiCom.2012} and a TB will be lost if the turbo-decoding processes require more than the available computational resources.

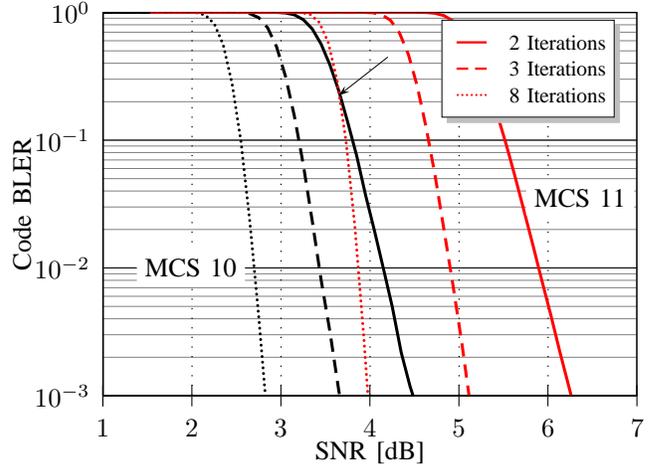
\begin{figure}[t]
\centering
\begingroup
\unitlength=1mm
\psset{xunit=11.83333mm, yunit=17.00000mm, linewidth=0.1mm}
\psset{arrowsize=2pt 3, arrowlength=1.4, arrowinset=.4}\psset{axesstyle=frame}
\begin{pspicture}(-0.18310, -3.82353)(7.00000, 0.00000)
\rput(-0.16901, -0.29412){%
\psaxes[subticks=10, ylogBase=10, logLines=y, labels=all, xsubticks=1, ysubticks=10, Ox=1, Oy=-3, Dx=1, Dy=1]{-}(1.00000, -3.00000)(1.00000, -3.00000)(7.00000, 0.00000)%
\multips(2.00000, -3.00000)(1.00000, 0.0){5}{\psline[linecolor=black, linestyle=dotted, linewidth=0.2mm](0, 0)(0, 3.00000)}
\rput[b](4.00000, -3.52941){SNR [dB]}
\rput[t]{90}(-0.01408, -1.50000){Code BLER}
\psclip{\psframe(1.00000, -3.00000)(7.00000, 0.00000)}
\psline[linecolor=black, plotstyle=curve, linewidth=0.4mm, showpoints=false, linestyle=dotted, dotsep=1pt, linecolor=black, dotstyle=square, dotscale=1.2 1.2, linewidth=0.4mm](0.94976, 0.00000)(1.04976, 0.00000)(1.14976, 0.00000)(1.24976, 0.00000)(1.34976, 0.00000)(1.44976, 0.00000)(1.54976, 0.00000)(1.64976, 0.00000)(1.74976, 0.00000)(1.84976, -0.00001)(1.94976, -0.00030)(2.04976, -0.00334)(2.14976, -0.02208)(2.24976, -0.09017)(2.34976, -0.25370)(2.44976, -0.55845)(2.54976, -1.01253)(2.64976, -1.64407)(2.74976, -2.38627)(2.84976, -3.22064)(2.94976, -3.71544)(3.04976, -3.96841)(3.14976, -4.23813)(3.24976, -4.28776)(3.34976, -4.70730)(3.44976, -4.45207)
\psline[linecolor=black, plotstyle=curve, linewidth=0.4mm, showpoints=false, linestyle=dashed, linecolor=black, dotstyle=square, dotscale=1.2 1.2, linewidth=0.4mm](0.94976, 0.00000)(1.04976, 0.00000)(1.14976, 0.00000)(1.24976, 0.00000)(1.34976, 0.00000)(1.44976, 0.00000)(1.54976, 0.00000)(1.64976, 0.00000)(1.74976, 0.00000)(1.84976, 0.00000)(1.94976, 0.00000)(2.04976, 0.00000)(2.14976, 0.00000)(2.24976, 0.00000)(2.34976, -0.00001)(2.44976, -0.00018)(2.54976, -0.00185)(2.64976, -0.01104)(2.74976, -0.04208)(2.84976, -0.12192)(2.94976, -0.27018)(3.04976, -0.50110)(3.14976, -0.81280)(3.24976, -1.19170)(3.34976, -1.62993)(3.44976, -2.09318)(3.54976, -2.50312)(3.64976, -2.96630)(3.74976, -3.33460)(3.84976, -3.93550)(3.94976, -4.00286)(4.04976, -4.77780)(4.14976, -4.28533)
\psline[linecolor=black, plotstyle=curve, linewidth=0.4mm, showpoints=false, linestyle=solid, linecolor=black, dotstyle=square, dotscale=1.2 1.2, linewidth=0.4mm](0.94976, 0.00000)(1.04976, 0.00000)(1.14976, 0.00000)(1.24976, 0.00000)(1.34976, 0.00000)(1.44976, 0.00000)(1.54976, 0.00000)(1.64976, 0.00000)(1.74976, 0.00000)(1.84976, 0.00000)(1.94976, 0.00000)(2.04976, 0.00000)(2.14976, 0.00000)(2.24976, 0.00000)(2.34976, 0.00000)(2.44976, 0.00000)(2.54976, 0.00000)(2.64976, 0.00000)(2.74976, -0.00001)(2.84976, -0.00013)(2.94976, -0.00100)(3.04976, -0.00518)(3.14976, -0.01989)(3.24976, -0.05724)(3.34976, -0.12930)(3.44976, -0.24809)(3.54976, -0.41310)(3.64976, -0.61689)(3.74976, -0.86208)(3.84976, -1.12325)(3.94976, -1.43135)(4.04976, -1.71073)(4.14976, -2.00582)(4.24976, -2.30591)(4.34976, -2.66253)(4.44976, -2.92888)(4.54976, -3.14560)(4.64976, -3.53522)(4.74976, -3.82832)(4.84976, -3.96559)(4.94976, -4.31681)(5.04976, -4.64673)(5.14976, -4.65993)(5.24976, -4.62605)(5.34976, -4.87436)(5.44976, -5.03632)
\psline[linecolor=red, plotstyle=curve, linewidth=0.4mm, showpoints=false, linestyle=dotted, dotsep=1pt, linecolor=red, dotstyle=square, dotscale=1.2 1.2, linewidth=0.4mm](1.52967, 0.00000)(1.62967, 0.00000)(1.72967, 0.00000)(1.82967, 0.00000)(1.92967, 0.00000)(2.02967, 0.00000)(2.12967, 0.00000)(2.22967, 0.00000)(2.32967, 0.00000)(2.42967, 0.00000)(2.52967, 0.00000)(2.62967, 0.00000)(2.72967, 0.00000)(2.82967, 0.00000)(2.92967, 0.00000)(3.02967, -0.00001)(3.12967, -0.00017)(3.22967, -0.00245)(3.32967, -0.01771)(3.42967, -0.07914)(3.52967, -0.23472)(3.62967, -0.53434)(3.72967, -1.00768)(3.82967, -1.67874)(3.92967, -2.54359)(4.02967, -3.52927)(4.12967, -4.57937)(4.22967, -4.75148)
\psline[linecolor=red, plotstyle=curve, linewidth=0.4mm, showpoints=false, linestyle=dashed, linecolor=red, dotstyle=square, dotscale=1.2 1.2, linewidth=0.4mm](1.52967, 0.00000)(1.62967, 0.00000)(1.72967, 0.00000)(1.82967, 0.00000)(1.92967, 0.00000)(2.02967, 0.00000)(2.12967, 0.00000)(2.22967, 0.00000)(2.32967, 0.00000)(2.42967, 0.00000)(2.52967, 0.00000)(2.62967, 0.00000)(2.72967, 0.00000)(2.82967, 0.00000)(2.92967, 0.00000)(3.02967, 0.00000)(3.12967, 0.00000)(3.22967, 0.00000)(3.32967, 0.00000)(3.42967, 0.00000)(3.52967, 0.00000)(3.62967, 0.00000)(3.72967, 0.00000)(3.82967, -0.00003)(3.92967, -0.00049)(4.02967, -0.00369)(4.12967, -0.01814)(4.22967, -0.06251)(4.32967, -0.16051)(4.42967, -0.32819)(4.52967, -0.57713)(4.62967, -0.90185)(4.72967, -1.27461)(4.82967, -1.69936)(4.92967, -2.12878)(5.02967, -2.58166)(5.12967, -3.09076)(5.22967, -3.41591)(5.32967, -3.81940)(5.42967, -4.06668)(5.52967, -4.51942)(5.62967, -4.84867)(5.72967, -5.21792)(5.82967, -4.98387)
\psline[linecolor=red, plotstyle=curve, linewidth=0.4mm, showpoints=false, linestyle=solid, linecolor=red, dotstyle=square, dotscale=1.2 1.2, linewidth=0.4mm](1.52967, 0.00000)(1.62967, 0.00000)(1.72967, 0.00000)(1.82967, 0.00000)(1.92967, 0.00000)(2.02967, 0.00000)(2.12967, 0.00000)(2.22967, 0.00000)(2.32967, 0.00000)(2.42967, 0.00000)(2.52967, 0.00000)(2.62967, 0.00000)(2.72967, 0.00000)(2.82967, 0.00000)(2.92967, 0.00000)(3.02967, 0.00000)(3.12967, 0.00000)(3.22967, 0.00000)(3.32967, 0.00000)(3.42967, 0.00000)(3.52967, 0.00000)(3.62967, 0.00000)(3.72967, 0.00000)(3.82967, 0.00000)(3.92967, 0.00000)(4.02967, 0.00000)(4.12967, 0.00000)(4.22967, 0.00000)(4.32967, -0.00000)(4.42967, -0.00001)(4.52967, -0.00029)(4.62967, -0.00161)(4.72967, -0.00778)(4.82967, -0.02575)(4.92967, -0.06642)(5.02967, -0.13907)(5.12967, -0.24974)(5.22967, -0.39896)(5.32967, -0.58582)(5.42967, -0.80084)(5.52967, -1.03490)(5.62967, -1.28767)(5.72967, -1.55408)(5.82967, -1.82190)(5.92967, -2.09896)(6.02967, -2.36651)(6.12967, -2.65086)(6.22967, -2.91944)(6.32967, -3.19104)(6.42967, -3.49346)(6.52967, -3.68557)(6.62967, -4.14810)(6.72967, -4.27915)(6.82967, -4.54224)(6.92967, -4.56579)(7.02967, -5.05469)(7.12967, -4.95739)(7.22967, -4.97259)
\endpsclip
\psline[linestyle=solid, linecolor=black]{->}(4.20000, -0.35000)(3.65000, -0.65000)
\rput[r](2.60000, -2.00000){\psframebox[linestyle=none, fillcolor=white, fillstyle=solid]{MCS 10}}
\rput[l](5.75000, -1.45000){\psframebox[linestyle=none, fillcolor=white, fillstyle=solid]{MCS 11}}
\psframe[linecolor=black, fillstyle=solid, fillcolor=white, shadowcolor=lightgray, shadowsize=1mm, shadow=true](4.80282, -0.82353)(6.74648, -0.05882)
\rput[l](5.56338, -0.23529){\footnotesize{2 Iterations}}
\psline[linecolor=red, linestyle=solid, linewidth=0.3mm](4.97183, -0.23529)(5.30986, -0.23529)
\rput[l](5.56338, -0.44118){\footnotesize{3 Iterations}}
\psline[linecolor=red, linestyle=dashed, linewidth=0.3mm](4.97183, -0.44118)(5.30986, -0.44118)
\rput[l](5.56338, -0.64706){\footnotesize{8 Iterations}}
\psline[linecolor=red, linestyle=dotted, dotsep=1pt, linewidth=0.3mm](4.97183, -0.64706)(5.30986, -0.64706)
}\end{pspicture}
\endgroup
 
\caption{Code block error rate as a function of SNR for MCS 10 and MCS 11 after 2, 3, and 8 decoder iterations.  The arrow shows where the CBLER of MCS 10 after 2 iterations is the same as that of MCS 11 after 8 iterations.}
\label{Figure:CBLER}
\end{figure}

The computational effort required to decode a turbo code is linear in the number of iterations and in the number of (information) bits.  Thus a reasonable metric for computational effort is the \emph{ bit-iteration}, defined as follows.   Let $I_r$ and $K_r$ be the number of executed iterations and the number of information bits associated with the $r^{th}$ CB of the TB, respectively.  In units of bit-iterations, the computational effort associated with the TB is
\vspace{-0.25cm}
\begin{eqnarray}
   \mathcal C (\gamma)
   & = &
   \sum_{r=1}^C
   K_r I_r
\end{eqnarray}
where the dependence of $K_r$ and $I_r$ on $\gamma$ is implicit.

The iterative nature of turbo decoding allows complexity to be traded off for transmission rate.  This is illustrated in Fig. \ref{Figure:CBLER}, which shows the \ac{CBLER} $\epsilon_\mathsf{cb}$ as a function of SNR (expressed as the ratio of the energy-per-symbol to noise power, $\mathcal E_s/N_0$) after 2, 3, and 8 iterations for two MCSs: $I_\mathsf{mcs} = 10$ and $11$.   The results in Fig.  \ref{Figure:CBLER} and the remaining examples in this paper assume that the UE  is allocated 45 \acp{RB}, which is the maximum allocation in a 10 MHz deployment when between three and five \acp{RB} are reserved for the physical uplink control channel \cite{3GPP.TS.36.211, 3GPP.TS.36.212}.
At $\epsilon_\mathsf{cb} = 2.5 \times 10^{-1}$, the performance of MCS 10 with 2 iterations is the same as the performance of MCS 11 with 8 iterations.  Thus, by backing off from MCS 11 to 10, the number of required iterations is cut by a factor of 4, on average.  The number of information bits in the TB that must be decoded is also reduced, in this case from 9216 to 8064, so the required number of bit-iterations is even further reduced.   However, this reduction in complexity comes at a cost, as the spectral efficiency of MCS 10 is lower than that of MCS 11.


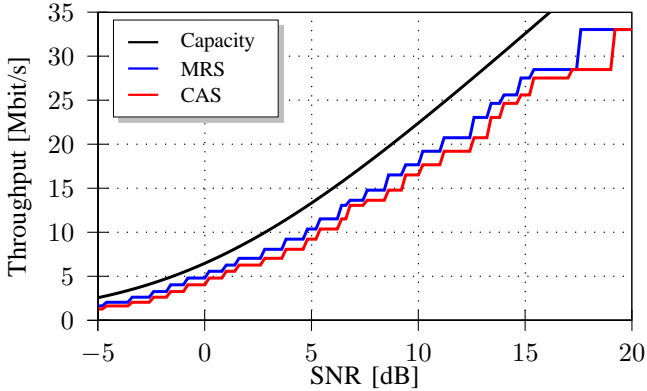
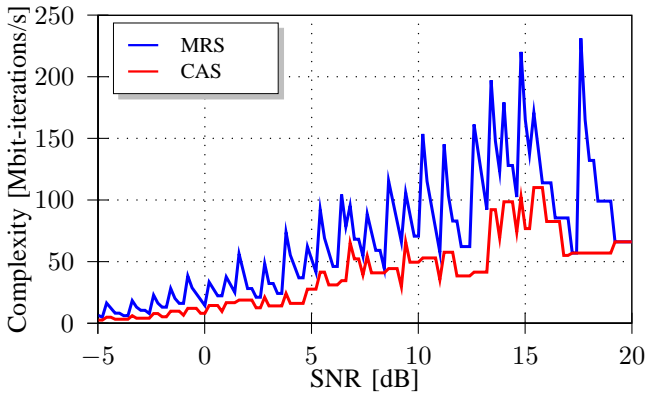
\begin{figure}
  \centering
  \subfigure[Raw throughput]{\begingroup
\unitlength=1mm
\psset{xunit=2.84000mm, yunit=1.17143mm, linewidth=0.1mm}
\psset{arrowsize=2pt 3, arrowlength=1.4, arrowinset=.4}\psset{axesstyle=frame}
\begin{pspicture}(-9.92958, -11.95122)(20.00000, 35.00000)
\rput(-0.70423, -4.26829){%
\psaxes[subticks=0, labels=all, xsubticks=1, ysubticks=1, Ox=-5, Oy=0, Dx=5, Dy=5]{-}(-5.00000, 0.00000)(-5.00000, 0.00000)(20.00000, 35.00000)%
\multips(0.00000, 0.00000)(5.00000, 0.0){4}{\psline[linecolor=black, linestyle=dotted, linewidth=0.2mm](0, 0)(0, 35.00000)}
\multips(-5.00000, 5.00000)(0, 5.00000){6}{\psline[linecolor=black, linestyle=dotted, linewidth=0.2mm](0, 0)(25.00000, 0)}
\rput[b](7.50000, -7.68293){SNR [dB]}
\rput[t]{90}(-9.22535, 17.50000){Throughput [Mbit/s]}
\psclip{\psframe(-5.00000, 0.00000)(20.00000, 35.00000)}
\psline[linecolor=blue, plotstyle=curve, linewidth=0.4mm, showpoints=false, linestyle=solid, linecolor=blue, dotstyle=triangle, dotscale=1.2 1.2, linewidth=0.4mm](-5.00000, 1.63200)(-4.80000, 1.63200)(-4.60000, 2.04800)(-4.40000, 2.04800)(-4.20000, 2.04800)(-4.00000, 2.04800)(-3.80000, 2.04800)(-3.60000, 2.04800)(-3.40000, 2.62400)(-3.20000, 2.62400)(-3.00000, 2.62400)(-2.80000, 2.62400)(-2.60000, 2.62400)(-2.40000, 3.26400)(-2.20000, 3.26400)(-2.00000, 3.26400)(-1.80000, 3.26400)(-1.60000, 4.03200)(-1.40000, 4.03200)(-1.20000, 4.03200)(-1.00000, 4.03200)(-0.80000, 4.80000)(-0.60000, 4.80000)(-0.40000, 4.80000)(-0.20000, 4.80000)(0.00000, 4.80000)(0.20000, 5.56800)(0.40000, 5.56800)(0.60000, 5.56800)(0.80000, 5.56800)(1.00000, 6.27200)(1.20000, 6.27200)(1.40000, 6.27200)(1.60000, 7.04000)(1.80000, 7.04000)(2.00000, 7.04000)(2.20000, 7.04000)(2.40000, 7.04000)(2.60000, 7.04000)(2.80000, 8.06400)(3.00000, 8.06400)(3.20000, 8.06400)(3.40000, 8.06400)(3.60000, 8.06400)(3.80000, 9.21600)(4.00000, 9.21600)(4.20000, 9.21600)(4.40000, 9.21600)(4.60000, 9.21600)(4.80000, 10.36800)(5.00000, 10.36800)(5.20000, 10.36800)(5.40000, 11.52000)(5.60000, 11.52000)(5.80000, 11.52000)(6.00000, 11.52000)(6.20000, 11.52000)(6.40000, 13.05600)(6.60000, 13.05600)(6.80000, 13.63200)(7.00000, 13.63200)(7.20000, 13.63200)(7.40000, 13.63200)(7.60000, 14.78400)(7.80000, 14.78400)(8.00000, 14.78400)(8.20000, 14.78400)(8.40000, 14.78400)(8.60000, 16.51200)(8.80000, 16.51200)(9.00000, 16.51200)(9.20000, 16.51200)(9.40000, 17.66400)(9.60000, 17.66400)(9.80000, 17.66400)(10.00000, 17.66400)(10.20000, 19.20000)(10.40000, 19.20000)(10.60000, 19.20000)(10.80000, 19.20000)(11.00000, 19.20000)(11.20000, 20.73600)(11.40000, 20.73600)(11.60000, 20.73600)(11.80000, 20.73600)(12.00000, 20.73600)(12.20000, 20.73600)(12.40000, 20.73600)(12.60000, 23.04000)(12.80000, 23.04000)(13.00000, 23.04000)(13.20000, 23.04000)(13.40000, 24.64000)(13.60000, 24.64000)(13.80000, 24.64000)(14.00000, 25.60000)(14.20000, 25.60000)(14.40000, 25.60000)(14.60000, 25.60000)(14.80000, 27.52000)(15.00000, 27.52000)(15.20000, 27.52000)(15.40000, 28.48000)(15.60000, 28.48000)(15.80000, 28.48000)(16.00000, 28.48000)(16.20000, 28.48000)(16.40000, 28.48000)(16.60000, 28.48000)(16.80000, 28.48000)(17.00000, 28.48000)(17.20000, 28.48000)(17.40000, 28.48000)(17.60000, 33.02400)(17.80000, 33.02400)(18.00000, 33.02400)(18.20000, 33.02400)(18.40000, 33.02400)(18.60000, 33.02400)(18.80000, 33.02400)(19.00000, 33.02400)(19.20000, 33.02400)(19.40000, 33.02400)(19.60000, 33.02400)(19.80000, 33.02400)(20.00000, 33.02400)
\psline[linecolor=red, plotstyle=curve, linewidth=0.4mm, showpoints=false, linestyle=solid, linecolor=red, dotstyle=x, dotscale=1.2 1.2, linewidth=0.4mm](-5.00000, 1.28000)(-4.80000, 1.28000)(-4.60000, 1.63200)(-4.40000, 1.63200)(-4.20000, 1.63200)(-4.00000, 1.63200)(-3.80000, 1.63200)(-3.60000, 1.63200)(-3.40000, 2.04800)(-3.20000, 2.04800)(-3.00000, 2.04800)(-2.80000, 2.04800)(-2.60000, 2.04800)(-2.40000, 2.62400)(-2.20000, 2.62400)(-2.00000, 2.62400)(-1.80000, 2.62400)(-1.60000, 3.26400)(-1.40000, 3.26400)(-1.20000, 3.26400)(-1.00000, 3.26400)(-0.80000, 4.03200)(-0.60000, 4.03200)(-0.40000, 4.03200)(-0.20000, 4.03200)(0.00000, 4.03200)(0.20000, 4.80000)(0.40000, 4.80000)(0.60000, 4.80000)(0.80000, 4.80000)(1.00000, 5.56800)(1.20000, 5.56800)(1.40000, 5.56800)(1.60000, 6.27200)(1.80000, 6.27200)(2.00000, 6.27200)(2.20000, 6.27200)(2.40000, 6.27200)(2.60000, 6.27200)(2.80000, 7.04000)(3.00000, 7.04000)(3.20000, 7.04000)(3.40000, 7.04000)(3.60000, 7.04000)(3.80000, 8.06400)(4.00000, 8.06400)(4.20000, 8.06400)(4.40000, 8.06400)(4.60000, 8.06400)(4.80000, 9.21600)(5.00000, 9.21600)(5.20000, 9.21600)(5.40000, 10.36800)(5.60000, 10.36800)(5.80000, 10.36800)(6.00000, 10.36800)(6.20000, 10.36800)(6.40000, 11.52000)(6.60000, 11.52000)(6.80000, 13.05600)(7.00000, 13.05600)(7.20000, 13.05600)(7.40000, 13.05600)(7.60000, 13.63200)(7.80000, 13.63200)(8.00000, 13.63200)(8.20000, 13.63200)(8.40000, 13.63200)(8.60000, 14.78400)(8.80000, 14.78400)(9.00000, 14.78400)(9.20000, 14.78400)(9.40000, 16.51200)(9.60000, 16.51200)(9.80000, 16.51200)(10.00000, 16.51200)(10.20000, 17.66400)(10.40000, 17.66400)(10.60000, 17.66400)(10.80000, 17.66400)(11.00000, 17.66400)(11.20000, 19.20000)(11.40000, 19.20000)(11.60000, 19.20000)(11.80000, 19.20000)(12.00000, 19.20000)(12.20000, 19.20000)(12.40000, 19.20000)(12.60000, 20.73600)(12.80000, 20.73600)(13.00000, 20.73600)(13.20000, 20.73600)(13.40000, 23.04000)(13.60000, 23.04000)(13.80000, 23.04000)(14.00000, 24.64000)(14.20000, 24.64000)(14.40000, 24.64000)(14.60000, 24.64000)(14.80000, 25.60000)(15.00000, 25.60000)(15.20000, 25.60000)(15.40000, 27.52000)(15.60000, 27.52000)(15.80000, 27.52000)(16.00000, 27.52000)(16.20000, 27.52000)(16.40000, 27.52000)(16.60000, 27.52000)(16.80000, 27.52000)(17.00000, 27.52000)(17.20000, 28.48000)(17.40000, 28.48000)(17.60000, 28.48000)(17.80000, 28.48000)(18.00000, 28.48000)(18.20000, 28.48000)(18.40000, 28.48000)(18.60000, 28.48000)(18.80000, 28.48000)(19.00000, 28.48000)(19.20000, 33.02400)(19.40000, 33.02400)(19.60000, 33.02400)(19.80000, 33.02400)(20.00000, 33.02400)
\psline[linecolor=black, plotstyle=curve, linewidth=0.4mm, showpoints=false, linestyle=solid, linecolor=black, dotstyle=square, dotscale=1.2 1.2, linewidth=0.4mm](-5.00000, 2.56873)(-4.80000, 2.67399)(-4.60000, 2.78295)(-4.40000, 2.89571)(-4.20000, 3.01233)(-4.00000, 3.13292)(-3.80000, 3.25754)(-3.60000, 3.38628)(-3.40000, 3.51921)(-3.20000, 3.65641)(-3.00000, 3.79795)(-2.80000, 3.94390)(-2.60000, 4.09433)(-2.40000, 4.24929)(-2.20000, 4.40885)(-2.00000, 4.57306)(-1.80000, 4.74197)(-1.60000, 4.91563)(-1.40000, 5.09409)(-1.20000, 5.27737)(-1.00000, 5.46552)(-0.80000, 5.65855)(-0.60000, 5.85650)(-0.40000, 6.05939)(-0.20000, 6.26722)(0.00000, 6.48000)(0.20000, 6.69774)(0.40000, 6.92043)(0.60000, 7.14807)(0.80000, 7.38064)(1.00000, 7.61813)(1.20000, 7.86050)(1.40000, 8.10774)(1.60000, 8.35981)(1.80000, 8.61667)(2.00000, 8.87828)(2.20000, 9.14459)(2.40000, 9.41555)(2.60000, 9.69111)(2.80000, 9.97121)(3.00000, 10.25578)(3.20000, 10.54476)(3.40000, 10.83808)(3.60000, 11.13567)(3.80000, 11.43746)(4.00000, 11.74336)(4.20000, 12.05329)(4.40000, 12.36719)(4.60000, 12.68496)(4.80000, 13.00651)(5.00000, 13.33178)(5.20000, 13.66066)(5.40000, 13.99308)(5.60000, 14.32893)(5.80000, 14.66815)(6.00000, 15.01064)(6.20000, 15.35630)(6.40000, 15.70506)(6.60000, 16.05683)(6.80000, 16.41152)(7.00000, 16.76904)(7.20000, 17.12931)(7.40000, 17.49224)(7.60000, 17.85776)(7.80000, 18.22578)(8.00000, 18.59622)(8.20000, 18.96900)(8.40000, 19.34405)(8.60000, 19.72128)(8.80000, 20.10063)(9.00000, 20.48201)(9.20000, 20.86537)(9.40000, 21.25062)(9.60000, 21.63771)(9.80000, 22.02656)(10.00000, 22.41712)(10.20000, 22.80931)(10.40000, 23.20308)(10.60000, 23.59837)(10.80000, 23.99512)(11.00000, 24.39328)(11.20000, 24.79278)(11.40000, 25.19359)(11.60000, 25.59565)(11.80000, 25.99890)(12.00000, 26.40331)(12.20000, 26.80883)(12.40000, 27.21541)(12.60000, 27.62301)(12.80000, 28.03158)(13.00000, 28.44110)(13.20000, 28.85152)(13.40000, 29.26280)(13.60000, 29.67492)(13.80000, 30.08782)(14.00000, 30.50149)(14.20000, 30.91589)(14.40000, 31.33099)(14.60000, 31.74676)(14.80000, 32.16317)(15.00000, 32.58019)(15.20000, 32.99781)(15.40000, 33.41599)(15.60000, 33.83471)(15.80000, 34.25395)(16.00000, 34.67368)(16.20000, 35.09388)(16.40000, 35.51454)(16.60000, 35.93564)(16.80000, 36.35715)(17.00000, 36.77905)(17.20000, 37.20134)(17.40000, 37.62399)(17.60000, 38.04699)(17.80000, 38.47032)(18.00000, 38.89397)(18.20000, 39.31793)(18.40000, 39.74218)(18.60000, 40.16670)(18.80000, 40.59149)(19.00000, 41.01654)(19.20000, 41.44183)(19.40000, 41.86735)(19.60000, 42.29309)(19.80000, 42.71905)(20.00000, 43.14521)
\endpsclip
\psframe[linecolor=black, fillstyle=solid, fillcolor=white, shadowcolor=lightgray, shadowsize=1mm, shadow=true](-4.29577, 23.04878)(3.45070, 34.14634)
\rput[l](-1.12676, 31.58537){\footnotesize{Capacity}}
\psline[linecolor=black, linestyle=solid, linewidth=0.3mm](-3.59155, 31.58537)(-2.18310, 31.58537)
\rput[l](-1.12676, 28.59756){\footnotesize{MRS}}
\psline[linecolor=blue, linestyle=solid, linewidth=0.3mm](-3.59155, 28.59756)(-2.18310, 28.59756)
\rput[l](-1.12676, 25.60976){\footnotesize{CAS}}
\psline[linecolor=red, linestyle=solid, linewidth=0.3mm](-3.59155, 25.60976)(-2.18310, 25.60976)
}\end{pspicture}
\endgroup
 }
  \subfigure[Computational effort]{\begingroup
\unitlength=1mm
\psset{xunit=2.84000mm, yunit=0.16400mm, linewidth=0.1mm}
\psset{arrowsize=2pt 3, arrowlength=1.4, arrowinset=.4}\psset{axesstyle=frame}
\begin{pspicture}(-9.92958, -85.36585)(20.00000, 250.00000)
\rput(-0.70423, -30.48780){%
\psaxes[subticks=0, labels=all, xsubticks=1, ysubticks=1, Ox=-5, Oy=0, Dx=5, Dy=50]{-}(-5.00000, 0.00000)(-5.00000, 0.00000)(20.00000, 250.00000)%
\multips(0.00000, 0.00000)(5.00000, 0.0){4}{\psline[linecolor=black, linestyle=dotted, linewidth=0.2mm](0, 0)(0, 250.00000)}
\multips(-5.00000, 50.00000)(0, 50.00000){4}{\psline[linecolor=black, linestyle=dotted, linewidth=0.2mm](0, 0)(25.00000, 0)}
\rput[b](7.50000, -54.87805){SNR [dB]}
\rput[t]{90}(-9.22535, 125.00000){Complexity [Mbit-iterations/s]}
\psclip{\psframe(-5.00000, 0.00000)(20.00000, 250.00000)}
\psline[linecolor=blue, plotstyle=curve, linewidth=0.4mm, showpoints=false, linestyle=solid, linecolor=blue, dotstyle=triangle, dotscale=1.2 1.2, linewidth=0.4mm](-5.00000, 6.52800)(-4.80000, 4.89600)(-4.60000, 16.38400)(-4.40000, 12.28800)(-4.20000, 8.19200)(-4.00000, 8.19200)(-3.80000, 6.14400)(-3.60000, 6.14400)(-3.40000, 18.36800)(-3.20000, 13.12000)(-3.00000, 10.49600)(-2.80000, 10.49600)(-2.60000, 7.87200)(-2.40000, 22.84800)(-2.20000, 16.32000)(-2.00000, 13.05600)(-1.80000, 13.05600)(-1.60000, 28.22400)(-1.40000, 20.16000)(-1.20000, 16.12800)(-1.00000, 16.12800)(-0.80000, 38.40000)(-0.60000, 28.80000)(-0.40000, 24.00000)(-0.20000, 19.20000)(0.00000, 14.40000)(0.20000, 33.40800)(0.40000, 27.84000)(0.60000, 22.27200)(0.80000, 22.27200)(1.00000, 37.63200)(1.20000, 31.36000)(1.40000, 25.08800)(1.60000, 56.32000)(1.80000, 42.24000)(2.00000, 28.16000)(2.20000, 28.16000)(2.40000, 21.12000)(2.60000, 21.12000)(2.80000, 48.38400)(3.00000, 32.25600)(3.20000, 32.25600)(3.40000, 24.19200)(3.60000, 24.19200)(3.80000, 73.72800)(4.00000, 55.29600)(4.20000, 46.08000)(4.40000, 36.86400)(4.60000, 36.86400)(4.80000, 62.20800)(5.00000, 51.84000)(5.20000, 41.47200)(5.40000, 92.16000)(5.60000, 69.12000)(5.80000, 57.60000)(6.00000, 46.08000)(6.20000, 46.08000)(6.40000, 104.44800)(6.60000, 78.33600)(6.80000, 95.42400)(7.00000, 68.16000)(7.20000, 68.16000)(7.40000, 54.52800)(7.60000, 88.70400)(7.80000, 73.92000)(8.00000, 59.13600)(8.20000, 59.13600)(8.40000, 44.35200)(8.60000, 115.58400)(8.80000, 99.07200)(9.00000, 82.56000)(9.20000, 66.04800)(9.40000, 105.98400)(9.60000, 88.32000)(9.80000, 70.65600)(10.00000, 70.65600)(10.20000, 153.60000)(10.40000, 115.20000)(10.60000, 96.00000)(10.80000, 76.80000)(11.00000, 57.60000)(11.20000, 145.15200)(11.40000, 103.68000)(11.60000, 82.94400)(11.80000, 82.94400)(12.00000, 62.20800)(12.20000, 62.20800)(12.40000, 62.20800)(12.60000, 161.28000)(12.80000, 138.24000)(13.00000, 115.20000)(13.20000, 92.16000)(13.40000, 197.12000)(13.60000, 147.84000)(13.80000, 123.20000)(14.00000, 179.20000)(14.20000, 128.00000)(14.40000, 128.00000)(14.60000, 102.40000)(14.80000, 220.16000)(15.00000, 165.12000)(15.20000, 137.60000)(15.40000, 170.88000)(15.60000, 142.40000)(15.80000, 113.92000)(16.00000, 113.92000)(16.20000, 113.92000)(16.40000, 85.44000)(16.60000, 85.44000)(16.80000, 85.44000)(17.00000, 85.44000)(17.20000, 56.96000)(17.40000, 56.96000)(17.60000, 231.16800)(17.80000, 165.12000)(18.00000, 132.09600)(18.20000, 132.09600)(18.40000, 99.07200)(18.60000, 99.07200)(18.80000, 99.07200)(19.00000, 99.07200)(19.20000, 66.04800)(19.40000, 66.04800)(19.60000, 66.04800)(19.80000, 66.04800)(20.00000, 66.04800)
\psline[linecolor=red, plotstyle=curve, linewidth=0.4mm, showpoints=false, linestyle=solid, linecolor=red, dotstyle=x, dotscale=1.2 1.2, linewidth=0.4mm](-5.00000, 2.56000)(-4.80000, 2.56000)(-4.60000, 4.89600)(-4.40000, 4.89600)(-4.20000, 3.26400)(-4.00000, 3.26400)(-3.80000, 3.26400)(-3.60000, 3.26400)(-3.40000, 6.14400)(-3.20000, 4.09600)(-3.00000, 4.09600)(-2.80000, 4.09600)(-2.60000, 4.09600)(-2.40000, 7.87200)(-2.20000, 7.87200)(-2.00000, 5.24800)(-1.80000, 5.24800)(-1.60000, 9.79200)(-1.40000, 9.79200)(-1.20000, 9.79200)(-1.00000, 6.52800)(-0.80000, 12.09600)(-0.60000, 12.09600)(-0.40000, 12.09600)(-0.20000, 8.06400)(0.00000, 8.06400)(0.20000, 14.40000)(0.40000, 14.40000)(0.60000, 14.40000)(0.80000, 9.60000)(1.00000, 16.70400)(1.20000, 16.70400)(1.40000, 16.70400)(1.60000, 18.81600)(1.80000, 18.81600)(2.00000, 18.81600)(2.20000, 18.81600)(2.40000, 12.54400)(2.60000, 12.54400)(2.80000, 21.12000)(3.00000, 14.08000)(3.20000, 14.08000)(3.40000, 14.08000)(3.60000, 14.08000)(3.80000, 24.19200)(4.00000, 16.12800)(4.20000, 16.12800)(4.40000, 16.12800)(4.60000, 16.12800)(4.80000, 27.64800)(5.00000, 27.64800)(5.20000, 27.64800)(5.40000, 41.47200)(5.60000, 41.47200)(5.80000, 31.10400)(6.00000, 31.10400)(6.20000, 31.10400)(6.40000, 34.56000)(6.60000, 34.56000)(6.80000, 65.28000)(7.00000, 52.22400)(7.20000, 52.22400)(7.40000, 39.16800)(7.60000, 54.52800)(7.80000, 40.89600)(8.00000, 40.89600)(8.20000, 40.89600)(8.40000, 40.89600)(8.60000, 44.35200)(8.80000, 44.35200)(9.00000, 44.35200)(9.20000, 29.56800)(9.40000, 66.04800)(9.60000, 49.53600)(9.80000, 49.53600)(10.00000, 49.53600)(10.20000, 52.99200)(10.40000, 52.99200)(10.60000, 52.99200)(10.80000, 52.99200)(11.00000, 35.32800)(11.20000, 57.60000)(11.40000, 57.60000)(11.60000, 57.60000)(11.80000, 38.40000)(12.00000, 38.40000)(12.20000, 38.40000)(12.40000, 38.40000)(12.60000, 41.47200)(12.80000, 41.47200)(13.00000, 41.47200)(13.20000, 41.47200)(13.40000, 92.16000)(13.60000, 92.16000)(13.80000, 69.12000)(14.00000, 98.56000)(14.20000, 98.56000)(14.40000, 98.56000)(14.60000, 73.92000)(14.80000, 102.40000)(15.00000, 76.80000)(15.20000, 76.80000)(15.40000, 110.08000)(15.60000, 110.08000)(15.80000, 110.08000)(16.00000, 82.56000)(16.20000, 82.56000)(16.40000, 82.56000)(16.60000, 82.56000)(16.80000, 55.04000)(17.00000, 55.04000)(17.20000, 56.96000)(17.40000, 56.96000)(17.60000, 56.96000)(17.80000, 56.96000)(18.00000, 56.96000)(18.20000, 56.96000)(18.40000, 56.96000)(18.60000, 56.96000)(18.80000, 56.96000)(19.00000, 56.96000)(19.20000, 66.04800)(19.40000, 66.04800)(19.60000, 66.04800)(19.80000, 66.04800)(20.00000, 66.04800)
\endpsclip
\psframe[linecolor=black, fillstyle=solid, fillcolor=white, shadowcolor=lightgray, shadowsize=1mm, shadow=true](-4.29577, 185.97561)(3.45070, 243.90244)
\rput[l](-1.12676, 225.60976){\footnotesize{MRS}}
\psline[linecolor=blue, linestyle=solid, linewidth=0.3mm](-3.59155, 225.60976)(-2.18310, 225.60976)
\rput[l](-1.12676, 204.26829){\footnotesize{CAS}}
\psline[linecolor=red, linestyle=solid, linewidth=0.3mm](-3.59155, 204.26829)(-2.18310, 204.26829)
}\end{pspicture}
\endgroup
 }
  \caption{Raw throughput and computational effort as a function of the SNR for two MCS selection schemes: computationally aware selection (CAS) and max-rate selection (MRS).
  }
  \label{fig:computation.spectral.eff.over.snr}
\end{figure}

The MCS is selected to satisfy a channel outage constraint $\hat{\epsilon}$.  In this paper, we use $\hat{\epsilon} = 0.1$, which is a typical value for an LTE network.  When complexity is not a concern, the selected MCS is the highest one that satisfies $\epsilon_\mathsf{channel} \leq \hat{\epsilon}$ after a large number of decoder iterations.   However, when complexity is a concern, the MCS should be selected such that the outage constraint is met after a specific number of decoder iterations.  Define $R_i(\gamma)$ to be the maximum rate for which $\epsilon_\mathsf{channel} \leq \hat{\epsilon}$ after the $i^{th}$ decoder iteration, where the maximization is over the set of MCSs.   The function $R_i(\gamma)$ determines the MCS-selection scheme.  When complexity is not a concern, $R_i(\gamma)$ with a large value of $i$ could be used, e.\,g. $i=8$.  In a complexity constrained deployment, using $R_i(\gamma)$ with a smaller value of $i$ could be used, where a sensible value might be $i=2$.     In the following, we refer to the \emph{\ac{MRS}} policy as the one that selects the MCS that achieves $R_8(\gamma)$, while we refer to the \emph{\ac{CAS}} policy as the one that selects the MCS that achieves $R_2(\gamma)$.  The minimum value of $\Delta \gamma$ for which $R_2( \gamma + \Delta \gamma) = R_8 ( \gamma )$ is an SNR margin required when CAS is used instead of MRS.


The rate-complexity tradeoff is illustrated in Fig. \ref{fig:computation.spectral.eff.over.snr}.   Fig. \ref{fig:computation.spectral.eff.over.snr}(a) shows the raw throughput of the MRS and CAS schemes as a function of the SNR \footnote{The \emph{raw} throughput $T_\mathsf{raw}$ is simply the selected rate $R_i(\gamma)$ expressed in units of bits/second.  This is in contrast with the \emph{effective} throughput, which is the rate of \emph{correct} transmission; i.e, $T_\mathsf{eff} = (1-\epsilon) T_\mathsf{raw}$.}.  For reference, the Shannon capacity limit is also shown.  The reduction in rate when using CAS is evident from the figure.  The average computational effort required to decode a TB as a function of SNR for the two MCS selection policies is shown in Fig. \ref{fig:computation.spectral.eff.over.snr}(b).   As anticipated, using the complexity aware scheme reduces the average effort.   The stairstep appearance of Fig. \ref{fig:computation.spectral.eff.over.snr}(a) and peaky behavior and Fig. \ref{fig:computation.spectral.eff.over.snr}(b) are due to the use of a finite number of MCSs.


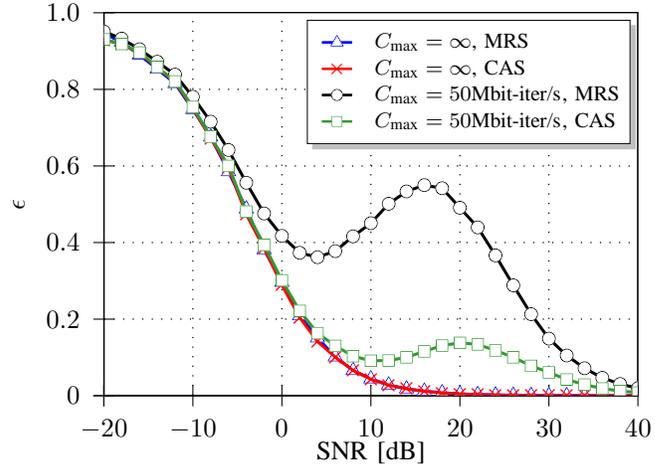
\begin{figure}[t]
\centering
\begingroup
\unitlength=1mm
\psset{xunit=1.18333mm, yunit=50.94905mm, linewidth=0.1mm}
\psset{arrowsize=2pt 3, arrowlength=1.4, arrowinset=.4}\psset{axesstyle=frame}
\begin{pspicture}(-31.83099, -0.27478)(40.00000, 1.00100)
\rput(-1.69014, -0.09814){%
\psaxes[subticks=0, labels=all, xsubticks=1, ysubticks=1, Ox=-20, Oy=0, Dx=10, Dy=0.2]{-}(-20.00000, 0.00000)(-20.00000, 0.00000)(40.00000, 1.00100)%
\multips(-10.00000, 0.00000)(10.00000, 0.0){5}{\psline[linecolor=black, linestyle=dotted, linewidth=0.2mm](0, 0)(0, 1.00100)}
\multips(-20.00000, 0.20000)(0, 0.20000){4}{\psline[linecolor=black, linestyle=dotted, linewidth=0.2mm](0, 0)(60.00000, 0)}
\rput[b](10.00000, -0.17665){SNR [dB]}
\rput[t]{90}(-30.14085, 0.50050){$\epsilon$}
\psclip{\psframe(-20.00000, 0.00000)(40.00000, 1.00100)}
\psline[linecolor=blue, plotstyle=curve, linewidth=0.4mm, showpoints=true, linestyle=solid, linecolor=blue, dotstyle=triangle, dotscale=1.2 1.2, linewidth=0.4mm](70.00000, 0.00000)(68.00000, 0.00000)(66.00000, 0.00000)(64.00000, 0.00000)(62.00000, 0.00000)(60.00000, 0.00000)(58.00000, 0.00000)(56.00000, 0.00000)(54.00000, 0.00000)(52.00000, 0.00000)(50.00000, 0.00001)(48.00000, 0.00001)(46.00000, 0.00002)(44.00000, 0.00002)(42.00000, 0.00004)(40.00000, 0.00006)(38.00000, 0.00009)(36.00000, 0.00014)(34.00000, 0.00022)(32.00000, 0.00035)(30.00000, 0.00052)(28.00000, 0.00087)(26.00000, 0.00136)(24.00000, 0.00203)(22.00000, 0.00318)(20.00000, 0.00471)(18.00000, 0.00783)(16.00000, 0.01181)(14.00000, 0.01785)(12.00000, 0.02822)(10.00000, 0.04417)(8.00000, 0.06594)(6.00000, 0.10147)(4.00000, 0.15115)(2.00000, 0.20888)(0.00000, 0.29635)(-2.00000, 0.38150)(-4.00000, 0.48795)(-6.00000, 0.58520)(-8.00000, 0.67601)(-10.00000, 0.74758)(-12.00000, 0.81546)(-14.00000, 0.85328)(-16.00000, 0.88992)(-18.00000, 0.92351)(-20.00000, 0.94620)(-22.00000, 0.95131)(-24.00000, 0.96310)(-26.00000, 0.97346)(-28.00000, 0.97883)(-30.00000, 0.98179)(-32.00000, 0.98810)(-34.00000, 0.99268)(-36.00000, 0.99286)(-38.00000, 0.99336)(-40.00000, 0.99552)(-42.00000, 0.99497)(-44.00000, 0.99678)(-46.00000, 0.99786)(-48.00000, 0.99874)(-50.00000, 0.99935)(-52.00000, 0.99941)(-54.00000, 0.99966)(-56.00000, 0.99899)(-58.00000, 0.99976)(-60.00000, 0.99947)
\psline[linecolor=red, plotstyle=curve, linewidth=0.4mm, showpoints=true, linestyle=solid, linecolor=red, dotstyle=x, dotscale=1.2 1.2, linewidth=0.4mm](70.00000, 0.00000)(68.00000, 0.00000)(66.00000, 0.00000)(64.00000, 0.00000)(62.00000, 0.00000)(60.00000, 0.00000)(58.00000, 0.00000)(56.00000, 0.00000)(54.00000, 0.00000)(52.00000, 0.00000)(50.00000, 0.00000)(48.00000, 0.00001)(46.00000, 0.00001)(44.00000, 0.00002)(42.00000, 0.00003)(40.00000, 0.00004)(38.00000, 0.00007)(36.00000, 0.00011)(34.00000, 0.00018)(32.00000, 0.00029)(30.00000, 0.00048)(28.00000, 0.00072)(26.00000, 0.00119)(24.00000, 0.00182)(22.00000, 0.00292)(20.00000, 0.00463)(18.00000, 0.00718)(16.00000, 0.01161)(14.00000, 0.01831)(12.00000, 0.02768)(10.00000, 0.04312)(8.00000, 0.06611)(6.00000, 0.10091)(4.00000, 0.14171)(2.00000, 0.20434)(0.00000, 0.28932)(-2.00000, 0.38515)(-4.00000, 0.47357)(-6.00000, 0.59478)(-8.00000, 0.67237)(-10.00000, 0.75097)(-12.00000, 0.81880)(-14.00000, 0.85703)(-16.00000, 0.89450)(-18.00000, 0.91765)(-20.00000, 0.92963)(-22.00000, 0.95309)(-24.00000, 0.96782)(-26.00000, 0.97059)(-28.00000, 0.97931)(-30.00000, 0.98362)(-32.00000, 0.98824)(-34.00000, 0.99080)(-36.00000, 0.99151)(-38.00000, 0.99658)(-40.00000, 0.99694)(-42.00000, 0.99719)(-44.00000, 0.99788)(-46.00000, 0.99900)(-48.00000, 0.99839)(-50.00000, 0.99870)(-52.00000, 0.99875)(-54.00000, 0.99994)(-56.00000, 0.99955)(-58.00000, 0.99997)(-60.00000, 0.99970)
\psline[linecolor=black, plotstyle=curve, linewidth=0.4mm, showpoints=true, linestyle=solid, linecolor=black, dotstyle=o, dotscale=1.2 1.2, linewidth=0.4mm](70.00000, 0.00002)(68.00000, 0.00003)(66.00000, 0.00005)(64.00000, 0.00008)(62.00000, 0.00013)(60.00000, 0.00020)(58.00000, 0.00032)(56.00000, 0.00052)(54.00000, 0.00079)(52.00000, 0.00129)(50.00000, 0.00202)(48.00000, 0.00318)(46.00000, 0.00515)(44.00000, 0.00796)(42.00000, 0.01301)(40.00000, 0.01968)(38.00000, 0.03052)(36.00000, 0.04673)(34.00000, 0.07173)(32.00000, 0.10512)(30.00000, 0.14914)(28.00000, 0.21294)(26.00000, 0.28821)(24.00000, 0.36645)(22.00000, 0.43958)(20.00000, 0.49087)(18.00000, 0.54187)(16.00000, 0.54993)(14.00000, 0.53346)(12.00000, 0.50123)(10.00000, 0.45056)(8.00000, 0.41618)(6.00000, 0.37756)(4.00000, 0.36177)(2.00000, 0.37319)(0.00000, 0.41734)(-2.00000, 0.47594)(-4.00000, 0.55619)(-6.00000, 0.64169)(-8.00000, 0.71558)(-10.00000, 0.78059)(-12.00000, 0.83853)(-14.00000, 0.87217)(-16.00000, 0.90480)(-18.00000, 0.93250)(-20.00000, 0.95110)(-22.00000, 0.95793)(-24.00000, 0.96698)(-26.00000, 0.97677)(-28.00000, 0.98153)(-30.00000, 0.98485)(-32.00000, 0.98932)(-34.00000, 0.99313)(-36.00000, 0.99377)(-38.00000, 0.99453)(-40.00000, 0.99607)(-42.00000, 0.99578)(-44.00000, 0.99706)(-46.00000, 0.99819)(-48.00000, 0.99879)(-50.00000, 0.99936)(-52.00000, 0.99966)(-54.00000, 0.99991)(-56.00000, 0.99900)(-58.00000, 0.99976)(-60.00000, 0.99951)
\psline[linecolor=darkgreen, plotstyle=curve, linewidth=0.4mm, showpoints=true, linestyle=solid, linecolor=darkgreen, dotstyle=square, dotscale=1.2 1.2, linewidth=0.4mm](70.00000, 0.00001)(68.00000, 0.00001)(66.00000, 0.00002)(64.00000, 0.00003)(62.00000, 0.00005)(60.00000, 0.00009)(58.00000, 0.00014)(56.00000, 0.00020)(54.00000, 0.00033)(52.00000, 0.00054)(50.00000, 0.00084)(48.00000, 0.00131)(46.00000, 0.00214)(44.00000, 0.00335)(42.00000, 0.00505)(40.00000, 0.00819)(38.00000, 0.01234)(36.00000, 0.01880)(34.00000, 0.02905)(32.00000, 0.04230)(30.00000, 0.06088)(28.00000, 0.07799)(26.00000, 0.10032)(24.00000, 0.11822)(22.00000, 0.13389)(20.00000, 0.13810)(18.00000, 0.13096)(16.00000, 0.11572)(14.00000, 0.10025)(12.00000, 0.09189)(10.00000, 0.09133)(8.00000, 0.10327)(6.00000, 0.13017)(4.00000, 0.16347)(2.00000, 0.22146)(0.00000, 0.30129)(-2.00000, 0.39403)(-4.00000, 0.48079)(-6.00000, 0.60007)(-8.00000, 0.67660)(-10.00000, 0.75406)(-12.00000, 0.82073)(-14.00000, 0.85868)(-16.00000, 0.89606)(-18.00000, 0.91857)(-20.00000, 0.93042)(-22.00000, 0.95347)(-24.00000, 0.96813)(-26.00000, 0.97083)(-28.00000, 0.97949)(-30.00000, 0.98387)(-32.00000, 0.98825)(-34.00000, 0.99096)(-36.00000, 0.99165)(-38.00000, 0.99658)(-40.00000, 0.99696)(-42.00000, 0.99724)(-44.00000, 0.99788)(-46.00000, 0.99900)(-48.00000, 0.99839)(-50.00000, 0.99870)(-52.00000, 0.99875)(-54.00000, 0.99994)(-56.00000, 0.99955)(-58.00000, 0.99997)(-60.00000, 0.99970)
\endpsclip
\psframe[linecolor=black, fillstyle=solid, fillcolor=white, shadowcolor=lightgray, shadowsize=1mm, shadow=true](2.81690, 0.65752)(39.15493, 0.98137)
\rput[l](10.42254, 0.92249){\footnotesize{$C_\text{max}=\infty$, MRS}}
\psline[linecolor=blue, linestyle=solid, linewidth=0.3mm](4.50704, 0.92249)(7.88732, 0.92249)
\psline[linecolor=blue, linestyle=solid, linewidth=0.3mm](4.50704, 0.92249)(7.88732, 0.92249)
\psdots[linecolor=blue, linestyle=solid, linewidth=0.3mm, dotstyle=triangle, dotscale=1.2 1.2, linecolor=blue](6.19718, 0.92249)
\rput[l](10.42254, 0.85379){\footnotesize{$C_\text{max}=\infty$, CAS}}
\psline[linecolor=red, linestyle=solid, linewidth=0.3mm](4.50704, 0.85379)(7.88732, 0.85379)
\psline[linecolor=red, linestyle=solid, linewidth=0.3mm](4.50704, 0.85379)(7.88732, 0.85379)
\psdots[linecolor=red, linestyle=solid, linewidth=0.3mm, dotstyle=x, dotscale=1.2 1.2, linecolor=red](6.19718, 0.85379)
\rput[l](10.42254, 0.78510){\footnotesize{$C_\text{max}=50\text{Mbit-iter/s}$, MRS}}
\psline[linecolor=black, linestyle=solid, linewidth=0.3mm](4.50704, 0.78510)(7.88732, 0.78510)
\psline[linecolor=black, linestyle=solid, linewidth=0.3mm](4.50704, 0.78510)(7.88732, 0.78510)
\psdots[linecolor=black, linestyle=solid, linewidth=0.3mm, dotstyle=o, dotscale=1.2 1.2, linecolor=black](6.19718, 0.78510)
\rput[l](10.42254, 0.71640){\footnotesize{$C_\text{max}=50\text{Mbit-iter/s}$, CAS}}
\psline[linecolor=darkgreen, linestyle=solid, linewidth=0.3mm](4.50704, 0.71640)(7.88732, 0.71640)
\psline[linecolor=darkgreen, linestyle=solid, linewidth=0.3mm](4.50704, 0.71640)(7.88732, 0.71640)
\psdots[linecolor=darkgreen, linestyle=solid, linewidth=0.3mm, dotstyle=square, dotscale=1.2 1.2, linecolor=darkgreen](6.19718, 0.71640)
}\end{pspicture}
\endgroup
 
\caption{Outage probability as a function of average SNR in the presence of Rayleigh fading, both with and without a complexity constraint.  }
 \label{Figure:OutageNoCCI}
\end{figure}

\subsection{Impact of Rayleigh Fading}

Now consider the complexity-throughput tradeoff in a fading channel.  When the fading is Rayleigh, the instantaneous SNR $\gamma$ is an exponential random variable with mean $\Gamma=\mathbb E [\gamma]$.  The complexity-outage tradeoff is obtained through  a simulation that works as follows.  During each trial, the SNR $\gamma$ is drawn from an exponential distribution.  The MCS scheme for the given $\gamma$ is determined according to the MCS selection policy.  Each of the $C$ \acp{CB} in the \ac{TB} is marked as being in a channel outage with probability $\epsilon_\mathsf{cb}$, which can be precomputed.  If any of the CB are in outage, then the entire TB is declared to be in an outage.  If the $r^{th}$ CB is in an outage, then $I_r = I_\mathsf{max}$, which is the maximum number of attempted iterations (typically 8).  Otherwise, $I_r$ is determined by drawing a random variable distributed according to the pdf of $I_r$, which can be precomputed by tracking the error-rate as a function of the number of iterations.



Fig. \ref{Figure:OutageNoCCI} shows the outage probability of both MCS-selection schemes in the presence of Rayleigh fading when there is no complexity constraint ($C_\mathsf{max}=\infty$) and when there is a complexity constraint of $C_\mathsf{max}=50$ Mbit-iterations per second\footnote{The complexity of a typical software turbo decoder requires between 100 and 1000 \ac{FLOPS} for each bit-iteration, depending on the actual implementation  \cite{Valenti.Sun.IJWIN.2001}.  For comparison, modern general-purpose processors support up to 50-200 billion \ac{FLOPS}.}.   As can be seen, using the more conservative CAS scheme greatly reduces the outage probability when complexity is constrained.  A non-monotonic behavior is observed in the complexity constrained outage curves.  At low SNR, channel outage dominates, as the conservative MCSs that must be selected do not have particularly high computational requirements due to having few information bits per TB.  However, as the average SNR increases, higher MCSs are selected more frequently and computational outage begins to dominate due to the additional computational burden associated with their larger payloads.  However, at very high average SNR, the instantaneous SNR is often much higher than the selection threshold for the highest MCS, and the CBs for that MCS will usually be successfully decoded after just one or two iterations, which is not enough load to trigger a computational outage.

\begin{figure}[t]
\centering
\begingroup
\unitlength=1mm
\psset{xunit=1.18333mm, yunit=1.45714mm, linewidth=0.1mm}
\psset{arrowsize=2pt 3, arrowlength=1.4, arrowinset=.4}\psset{axesstyle=frame}
\begin{pspicture}(-31.83099, -9.60785)(40.00000, 35.00001)
\rput(-1.69014, -3.43137){%
\psaxes[subticks=0, labels=all, xsubticks=1, ysubticks=1, Ox=-20, Oy=0, Dx=10, Dy=5]{-}(-20.00000, 0.00000)(-20.00000, 0.00000)(40.00000, 35.00001)%
\multips(-10.00000, 0.00000)(10.00000, 0.0){5}{\psline[linecolor=black, linestyle=dotted, linewidth=0.2mm](0, 0)(0, 35.00001)}
\multips(-20.00000, 5.00000)(0, 5.00000){6}{\psline[linecolor=black, linestyle=dotted, linewidth=0.2mm](0, 0)(60.00000, 0)}
\rput[b](10.00000, -6.17647){SNR [dB]}
\rput[t]{90}(-30.14085, 17.50001){Throughput [MBit/s]}
\psclip{\psframe(-20.00000, 0.00000)(40.00000, 35.00001)}
\psline[linecolor=blue, plotstyle=curve, linewidth=0.4mm, showpoints=true, linestyle=solid, linecolor=blue, dotstyle=triangle, dotscale=1.2 1.2, linewidth=0.4mm](70.00000, 33.02387)(68.00000, 33.02380)(66.00000, 33.02368)(64.00000, 33.02349)(62.00000, 33.02322)(60.00000, 33.02278)(58.00000, 33.02208)(56.00000, 33.02078)(54.00000, 33.01914)(52.00000, 33.01610)(50.00000, 33.01176)(48.00000, 33.00468)(46.00000, 32.99241)(44.00000, 32.97526)(42.00000, 32.94371)(40.00000, 32.90190)(38.00000, 32.83240)(36.00000, 32.72560)(34.00000, 32.55354)(32.00000, 32.30203)(30.00000, 31.95163)(28.00000, 31.34160)(26.00000, 30.53725)(24.00000, 29.52816)(22.00000, 28.19088)(20.00000, 26.75680)(18.00000, 24.41425)(16.00000, 22.34306)(14.00000, 20.31521)(12.00000, 17.83965)(10.00000, 15.58374)(8.00000, 13.25379)(6.00000, 11.06299)(4.00000, 9.08845)(2.00000, 7.50889)(0.00000, 5.77328)(-2.00000, 4.67756)(-4.00000, 3.57307)(-6.00000, 2.95554)(-8.00000, 2.21516)(-10.00000, 1.67939)(-12.00000, 1.25026)(-14.00000, 0.98994)(-16.00000, 0.74578)(-18.00000, 0.50149)(-20.00000, 0.32037)(-22.00000, 0.33154)(-24.00000, 0.24373)(-26.00000, 0.17759)(-28.00000, 0.13548)(-30.00000, 0.13064)(-32.00000, 0.06314)(-34.00000, 0.03228)(-36.00000, 0.04254)(-38.00000, 0.05209)(-40.00000, 0.03532)(-42.00000, 0.03695)(-44.00000, 0.01588)(-46.00000, 0.01401)(-48.00000, 0.00471)(-50.00000, 0.00219)(-52.00000, 0.00859)(-54.00000, 0.00681)(-56.00000, 0.00324)(-58.00000, 0.00053)(-60.00000, 0.00280)
\psline[linecolor=red, plotstyle=curve, linewidth=0.4mm, showpoints=true, linestyle=solid, linecolor=red, dotstyle=x, dotscale=1.2 1.2, linewidth=0.4mm](70.00000, 33.02380)(68.00000, 33.02373)(66.00000, 33.02353)(64.00000, 33.02329)(62.00000, 33.02282)(60.00000, 33.02217)(58.00000, 33.02099)(56.00000, 33.01963)(54.00000, 33.01682)(52.00000, 33.01205)(50.00000, 33.00587)(48.00000, 32.99539)(46.00000, 32.97689)(44.00000, 32.94985)(42.00000, 32.91241)(40.00000, 32.84202)(38.00000, 32.74459)(36.00000, 32.58993)(34.00000, 32.33072)(32.00000, 31.96206)(30.00000, 31.37069)(28.00000, 30.72379)(26.00000, 29.62558)(24.00000, 28.40423)(22.00000, 26.66897)(20.00000, 24.80272)(18.00000, 22.71410)(16.00000, 20.30905)(14.00000, 17.88539)(12.00000, 15.85948)(10.00000, 13.63825)(8.00000, 11.57582)(6.00000, 9.75645)(4.00000, 8.14533)(2.00000, 6.52862)(0.00000, 5.13823)(-2.00000, 3.98379)(-4.00000, 3.24859)(-6.00000, 2.30536)(-8.00000, 1.84963)(-10.00000, 1.36871)(-12.00000, 1.02352)(-14.00000, 0.79352)(-16.00000, 0.61149)(-18.00000, 0.45996)(-20.00000, 0.39798)(-22.00000, 0.25726)(-24.00000, 0.16554)(-26.00000, 0.15140)(-28.00000, 0.11536)(-30.00000, 0.09951)(-32.00000, 0.04923)(-34.00000, 0.05555)(-36.00000, 0.04219)(-38.00000, 0.01194)(-40.00000, 0.01333)(-42.00000, 0.02354)(-44.00000, 0.00670)(-46.00000, 0.00285)(-48.00000, 0.00593)(-50.00000, 0.00708)(-52.00000, 0.00440)(-54.00000, 0.00008)(-56.00000, 0.00107)(-58.00000, 0.00004)(-60.00000, 0.00052)
\psline[linecolor=black, plotstyle=curve, linewidth=0.4mm, showpoints=true, linestyle=solid, linecolor=black, dotstyle=o, dotscale=1.2 1.2, linewidth=0.4mm](70.00000, 33.02329)(68.00000, 33.02287)(66.00000, 33.02222)(64.00000, 33.02114)(62.00000, 33.01951)(60.00000, 33.01701)(58.00000, 33.01284)(56.00000, 33.00593)(54.00000, 32.99643)(52.00000, 32.97924)(50.00000, 32.95406)(48.00000, 32.91358)(46.00000, 32.84516)(44.00000, 32.74759)(42.00000, 32.57197)(40.00000, 32.33980)(38.00000, 31.96212)(36.00000, 31.39616)(34.00000, 30.52067)(32.00000, 29.34113)(30.00000, 27.77780)(28.00000, 25.46435)(26.00000, 22.69360)(24.00000, 19.71830)(22.00000, 16.71186)(20.00000, 14.29216)(18.00000, 11.28234)(16.00000, 9.63236)(14.00000, 8.57690)(12.00000, 7.57654)(10.00000, 6.87701)(8.00000, 6.13406)(6.00000, 5.56479)(4.00000, 4.96999)(2.00000, 4.31518)(0.00000, 3.41523)(-2.00000, 2.85143)(-4.00000, 2.24514)(-6.00000, 1.83135)(-8.00000, 1.44443)(-10.00000, 1.04391)(-12.00000, 0.78870)(-14.00000, 0.62261)(-16.00000, 0.46390)(-18.00000, 0.32577)(-20.00000, 0.22932)(-22.00000, 0.20526)(-24.00000, 0.16796)(-26.00000, 0.11147)(-28.00000, 0.08499)(-30.00000, 0.06831)(-32.00000, 0.04197)(-34.00000, 0.02472)(-36.00000, 0.02641)(-38.00000, 0.02994)(-40.00000, 0.02405)(-42.00000, 0.02180)(-44.00000, 0.01138)(-46.00000, 0.00770)(-48.00000, 0.00400)(-50.00000, 0.00206)(-52.00000, 0.00230)(-54.00000, 0.00112)(-56.00000, 0.00313)(-58.00000, 0.00053)(-60.00000, 0.00233)
\psline[linecolor=darkgreen, plotstyle=curve, linewidth=0.4mm, showpoints=true, linestyle=solid, linecolor=darkgreen, dotstyle=square, dotscale=1.2 1.2, linewidth=0.4mm](70.00000, 33.02353)(68.00000, 33.02338)(66.00000, 33.02285)(64.00000, 33.02228)(62.00000, 33.02123)(60.00000, 33.01967)(58.00000, 33.01693)(56.00000, 33.01371)(54.00000, 33.00714)(52.00000, 32.99616)(50.00000, 32.98117)(48.00000, 32.95689)(46.00000, 32.91393)(44.00000, 32.85135)(42.00000, 32.76397)(40.00000, 32.60134)(38.00000, 32.38254)(36.00000, 32.03906)(34.00000, 31.48069)(32.00000, 30.72785)(30.00000, 29.60198)(28.00000, 28.46737)(26.00000, 26.74345)(24.00000, 25.03292)(22.00000, 22.89675)(20.00000, 20.97749)(18.00000, 19.18813)(16.00000, 17.35865)(14.00000, 15.56819)(12.00000, 14.04425)(10.00000, 12.27601)(8.00000, 10.52636)(6.00000, 8.92906)(4.00000, 7.52972)(2.00000, 6.04595)(0.00000, 4.79899)(-2.00000, 3.73297)(-4.00000, 3.04409)(-6.00000, 2.15592)(-8.00000, 1.73020)(-10.00000, 1.28149)(-12.00000, 0.96901)(-14.00000, 0.74709)(-16.00000, 0.56749)(-18.00000, 0.43394)(-20.00000, 0.37578)(-22.00000, 0.24625)(-24.00000, 0.15675)(-26.00000, 0.14458)(-28.00000, 0.11035)(-30.00000, 0.09225)(-32.00000, 0.04891)(-34.00000, 0.05098)(-36.00000, 0.03817)(-38.00000, 0.01194)(-40.00000, 0.01284)(-42.00000, 0.02204)(-44.00000, 0.00670)(-46.00000, 0.00285)(-48.00000, 0.00593)(-50.00000, 0.00705)(-52.00000, 0.00440)(-54.00000, 0.00008)(-56.00000, 0.00107)(-58.00000, 0.00004)(-60.00000, 0.00052)
\endpsclip
\psframe[linecolor=black, fillstyle=solid, fillcolor=white, shadowcolor=lightgray, shadowsize=1mm, shadow=true](-19.15493, 25.04903)(17.18310, 36.37256)
\rput[l](-11.54930, 34.31374){\footnotesize{$C_\text{max}=\infty$, MRS}}
\psline[linecolor=blue, linestyle=solid, linewidth=0.3mm](-17.46479, 34.31374)(-14.08451, 34.31374)
\psline[linecolor=blue, linestyle=solid, linewidth=0.3mm](-17.46479, 34.31374)(-14.08451, 34.31374)
\psdots[linecolor=blue, linestyle=solid, linewidth=0.3mm, dotstyle=triangle, dotscale=1.2 1.2, linecolor=blue](-15.77465, 34.31374)
\rput[l](-11.54930, 31.91177){\footnotesize{$C_\text{max}=\infty$, CAS}}
\psline[linecolor=red, linestyle=solid, linewidth=0.3mm](-17.46479, 31.91177)(-14.08451, 31.91177)
\psline[linecolor=red, linestyle=solid, linewidth=0.3mm](-17.46479, 31.91177)(-14.08451, 31.91177)
\psdots[linecolor=red, linestyle=solid, linewidth=0.3mm, dotstyle=x, dotscale=1.2 1.2, linecolor=red](-15.77465, 31.91177)
\rput[l](-11.54930, 29.50981){\footnotesize{$C_\text{max}=50\text{Mbit-iter/s}$, MRS}}
\psline[linecolor=black, linestyle=solid, linewidth=0.3mm](-17.46479, 29.50981)(-14.08451, 29.50981)
\psline[linecolor=black, linestyle=solid, linewidth=0.3mm](-17.46479, 29.50981)(-14.08451, 29.50981)
\psdots[linecolor=black, linestyle=solid, linewidth=0.3mm, dotstyle=o, dotscale=1.2 1.2, linecolor=black](-15.77465, 29.50981)
\rput[l](-11.54930, 27.10785){\footnotesize{$C_\text{max}=50\text{Mbit-iter/s}$, CAS}}
\psline[linecolor=darkgreen, linestyle=solid, linewidth=0.3mm](-17.46479, 27.10785)(-14.08451, 27.10785)
\psline[linecolor=darkgreen, linestyle=solid, linewidth=0.3mm](-17.46479, 27.10785)(-14.08451, 27.10785)
\psdots[linecolor=darkgreen, linestyle=solid, linewidth=0.3mm, dotstyle=square, dotscale=1.2 1.2, linecolor=darkgreen](-15.77465, 27.10785)
}\end{pspicture}
\endgroup
 
\caption{Effective throughput as a function of average SNR in the presence of Rayleigh fading, both with and without a complexity constraint.   }
 \label{Figure:ThroughputNoCCI}
\end{figure}
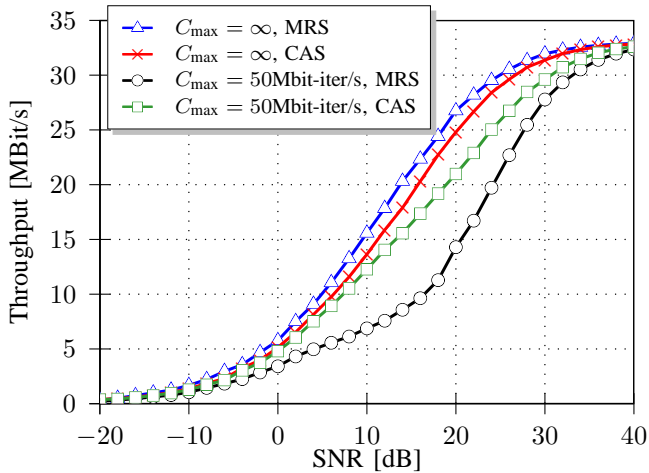

In a complexity constrained system, the advantage of a complexity aware MCS selection is a reduction in outage probability, but its disadvantage is a loss in the raw throughput.    However, if the outage probability is sufficiently lower, the effective throughput may be higher despite the reduced raw throughput.  This can be seen in Fig. \ref{Figure:ThroughputNoCCI}, which shows the effective throughput as a function of average SNR for the same cases that were shown in Fig. \ref{Figure:OutageNoCCI}.   The loss in effective throughput due to the complexity constraint is evident in the figure, but much of this loss can be recovered by using complexity aware MCS selection; i.e., the throughput curve for CAS is up to 9 dB higher than the curve for MRS.

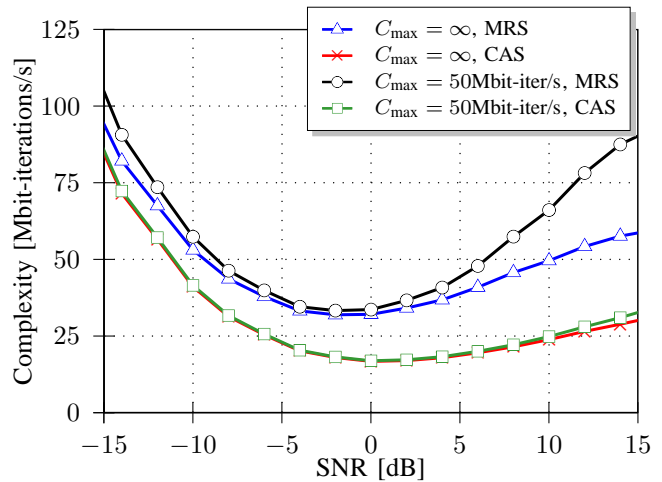
\begin{figure}[t]
\centering
\begingroup
\unitlength=1mm
\psset{xunit=2.36667mm, yunit=0.40800mm, linewidth=0.1mm}
\psset{arrowsize=2pt 3, arrowlength=1.4, arrowinset=.4}\psset{axesstyle=frame}
\begin{pspicture}(-20.91549, -34.31373)(15.00000, 125.00001)
\rput(-0.84507, -12.25490){%
\psaxes[subticks=0, labels=all, xsubticks=1, ysubticks=1, Ox=-15, Oy=0, Dx=5, Dy=25]{-}(-15.00000, 0.00000)(-15.00000, 0.00000)(15.00000, 125.00001)%
\multips(-10.00000, 0.00000)(5.00000, 0.0){5}{\psline[linecolor=black, linestyle=dotted, linewidth=0.2mm](0, 0)(0, 125.00001)}
\multips(-15.00000, 25.00000)(0, 25.00000){4}{\psline[linecolor=black, linestyle=dotted, linewidth=0.2mm](0, 0)(30.00000, 0)}
\rput[b](0.00000, -22.05883){SNR [dB]}
\rput[t]{90}(-20.07042, 62.50001){Complexity [Mbit-iterations/s]}
\psclip{\psframe(-15.00000, 0.00000)(15.00000, 125.00001)}
\psline[linecolor=blue, plotstyle=curve, linewidth=0.4mm, showpoints=true, linestyle=solid, linecolor=blue, dotstyle=triangle, dotscale=1.2 1.2, linewidth=0.4mm](70.00000, 33.02505)(68.00000, 33.02566)(66.00000, 33.02658)(64.00000, 33.02829)(62.00000, 33.03057)(60.00000, 33.03428)(58.00000, 33.04051)(56.00000, 33.05069)(54.00000, 33.06463)(52.00000, 33.09017)(50.00000, 33.12756)(48.00000, 33.18789)(46.00000, 33.28889)(44.00000, 33.43325)(42.00000, 33.69342)(40.00000, 34.03893)(38.00000, 34.59676)(36.00000, 35.43412)(34.00000, 36.72767)(32.00000, 38.45672)(30.00000, 40.74887)(28.00000, 44.07039)(26.00000, 47.99943)(24.00000, 52.05069)(22.00000, 55.71923)(20.00000, 58.14112)(18.00000, 60.18941)(16.00000, 59.69235)(14.00000, 57.59088)(12.00000, 54.20080)(10.00000, 49.57989)(8.00000, 45.70644)(6.00000, 40.90450)(4.00000, 36.80582)(2.00000, 34.13269)(0.00000, 32.08318)(-2.00000, 31.93101)(-4.00000, 33.19511)(-6.00000, 37.88217)(-8.00000, 43.61498)(-10.00000, 53.05402)(-12.00000, 67.50388)(-14.00000, 82.09828)(-16.00000, 106.02204)(-18.00000, 145.23536)(-20.00000, 199.66127)(-22.00000, 223.36578)(-24.00000, 287.96998)(-26.00000, 397.81237)(-28.00000, 496.14553)(-30.00000, 577.46748)
\psline[linecolor=red, plotstyle=curve, linewidth=0.4mm, showpoints=true, linestyle=solid, linecolor=red, dotstyle=x, dotscale=1.2 1.2, linewidth=0.4mm](70.00000, 33.02427)(68.00000, 33.02434)(66.00000, 33.02467)(64.00000, 33.02500)(62.00000, 33.02558)(60.00000, 33.02652)(58.00000, 33.02809)(56.00000, 33.03002)(54.00000, 33.03375)(52.00000, 33.03995)(50.00000, 33.04887)(48.00000, 33.06295)(46.00000, 33.08737)(44.00000, 33.12314)(42.00000, 33.17316)(40.00000, 33.26483)(38.00000, 33.38469)(36.00000, 33.56984)(34.00000, 33.85634)(32.00000, 34.21253)(30.00000, 34.68205)(28.00000, 35.07445)(26.00000, 35.48364)(24.00000, 35.65966)(22.00000, 35.51638)(20.00000, 34.79420)(18.00000, 33.43746)(16.00000, 31.33226)(14.00000, 28.85751)(12.00000, 26.51870)(10.00000, 23.84351)(8.00000, 21.49090)(6.00000, 19.54473)(4.00000, 17.93386)(2.00000, 16.99712)(0.00000, 16.73643)(-2.00000, 18.00594)(-4.00000, 20.16737)(-6.00000, 25.38459)(-8.00000, 31.35133)(-10.00000, 41.05703)(-12.00000, 56.56691)(-14.00000, 71.51384)(-16.00000, 97.42212)(-18.00000, 124.26338)(-20.00000, 145.82439)(-22.00000, 218.00124)(-24.00000, 317.63554)(-26.00000, 347.26216)(-28.00000, 494.92132)(-30.00000, 625.44624)
\psline[linecolor=black, plotstyle=curve, linewidth=0.4mm, showpoints=true, linestyle=solid, linecolor=black, dotstyle=o, dotscale=1.2 1.2, linewidth=0.4mm](70.00000, 33.02508)(68.00000, 33.02570)(66.00000, 33.02668)(64.00000, 33.02832)(62.00000, 33.03075)(60.00000, 33.03456)(58.00000, 33.04088)(56.00000, 33.05119)(54.00000, 33.06563)(52.00000, 33.09158)(50.00000, 33.13000)(48.00000, 33.19121)(46.00000, 33.29502)(44.00000, 33.44424)(42.00000, 33.71404)(40.00000, 34.07531)(38.00000, 34.67097)(36.00000, 35.58557)(34.00000, 37.05560)(32.00000, 39.13586)(30.00000, 42.12191)(28.00000, 47.00921)(26.00000, 53.87673)(24.00000, 62.68669)(22.00000, 73.00965)(20.00000, 81.81613)(18.00000, 92.09858)(16.00000, 93.00110)(14.00000, 87.47464)(12.00000, 78.16652)(10.00000, 66.08334)(8.00000, 57.41972)(6.00000, 47.85849)(4.00000, 40.81410)(2.00000, 36.63796)(0.00000, 33.63167)(-2.00000, 33.29894)(-4.00000, 34.55616)(-6.00000, 39.88193)(-8.00000, 46.30587)(-10.00000, 57.36591)(-12.00000, 73.54323)(-14.00000, 90.61588)(-16.00000, 118.87162)(-18.00000, 161.32633)(-20.00000, 217.35428)(-22.00000, 254.84697)(-24.00000, 319.10054)(-26.00000, 450.98011)(-28.00000, 565.23687)(-30.00000, 688.69428)
\psline[linecolor=darkgreen, plotstyle=curve, linewidth=0.4mm, showpoints=true, linestyle=solid, linecolor=darkgreen, dotstyle=square, dotscale=1.2 1.2, linewidth=0.4mm](70.00000, 33.02450)(68.00000, 33.02465)(66.00000, 33.02528)(64.00000, 33.02590)(62.00000, 33.02699)(60.00000, 33.02874)(58.00000, 33.03169)(56.00000, 33.03524)(54.00000, 33.04234)(52.00000, 33.05402)(50.00000, 33.07081)(48.00000, 33.09710)(46.00000, 33.14334)(44.00000, 33.21107)(42.00000, 33.30626)(40.00000, 33.48200)(38.00000, 33.71508)(36.00000, 34.08070)(34.00000, 34.66368)(32.00000, 35.42303)(30.00000, 36.49323)(28.00000, 37.47616)(26.00000, 38.70255)(24.00000, 39.55621)(22.00000, 39.98661)(20.00000, 39.28651)(18.00000, 37.40542)(16.00000, 34.38519)(14.00000, 31.00010)(12.00000, 28.01754)(10.00000, 24.82715)(8.00000, 22.16247)(6.00000, 20.02313)(4.00000, 18.26157)(2.00000, 17.25578)(0.00000, 16.93014)(-2.00000, 18.19214)(-4.00000, 20.37320)(-6.00000, 25.65075)(-8.00000, 31.69363)(-10.00000, 41.50797)(-12.00000, 57.12338)(-14.00000, 72.28667)(-16.00000, 98.80280)(-18.00000, 125.61445)(-20.00000, 147.42909)(-22.00000, 219.76223)(-24.00000, 320.73103)(-26.00000, 350.11042)(-28.00000, 499.19857)(-30.00000, 635.09211)
\endpsclip
\psframe[linecolor=black, fillstyle=solid, fillcolor=white, shadowcolor=lightgray, shadowsize=1mm, shadow=true](-3.59155, 91.91177)(14.57746, 132.35295)
\rput[l](0.21127, 125.00001){\footnotesize{$C_\text{max}=\infty$, MRS}}
\psline[linecolor=blue, linestyle=solid, linewidth=0.3mm](-2.74648, 125.00001)(-1.05634, 125.00001)
\psline[linecolor=blue, linestyle=solid, linewidth=0.3mm](-2.74648, 125.00001)(-1.05634, 125.00001)
\psdots[linecolor=blue, linestyle=solid, linewidth=0.3mm, dotstyle=triangle, dotscale=1.2 1.2, linecolor=blue](-1.90141, 125.00001)
\rput[l](0.21127, 116.42158){\footnotesize{$C_\text{max}=\infty$, CAS}}
\psline[linecolor=red, linestyle=solid, linewidth=0.3mm](-2.74648, 116.42158)(-1.05634, 116.42158)
\psline[linecolor=red, linestyle=solid, linewidth=0.3mm](-2.74648, 116.42158)(-1.05634, 116.42158)
\psdots[linecolor=red, linestyle=solid, linewidth=0.3mm, dotstyle=x, dotscale=1.2 1.2, linecolor=red](-1.90141, 116.42158)
\rput[l](0.21127, 107.84315){\footnotesize{$C_\text{max}=50\text{Mbit-iter/s}$, MRS}}
\psline[linecolor=black, linestyle=solid, linewidth=0.3mm](-2.74648, 107.84315)(-1.05634, 107.84315)
\psline[linecolor=black, linestyle=solid, linewidth=0.3mm](-2.74648, 107.84315)(-1.05634, 107.84315)
\psdots[linecolor=black, linestyle=solid, linewidth=0.3mm, dotstyle=o, dotscale=1.2 1.2, linecolor=black](-1.90141, 107.84315)
\rput[l](0.21127, 99.26471){\footnotesize{$C_\text{max}=50\text{Mbit-iter/s}$, CAS}}
\psline[linecolor=darkgreen, linestyle=solid, linewidth=0.3mm](-2.74648, 99.26471)(-1.05634, 99.26471)
\psline[linecolor=darkgreen, linestyle=solid, linewidth=0.3mm](-2.74648, 99.26471)(-1.05634, 99.26471)
\psdots[linecolor=darkgreen, linestyle=solid, linewidth=0.3mm, dotstyle=square, dotscale=1.2 1.2, linecolor=darkgreen](-1.90141, 99.26471)
}\end{pspicture}
\endgroup
 
\caption{Average complexity required in Rayleigh fading for a successfully decoded codeword as function of average SNR.  }
 \label{Figure:ComplexitySDC}
\end{figure}



The average complexity required for successful decoding in Rayleigh fading is shown in Fig. \ref{Figure:ComplexitySDC}.  The MRS scheme always has a higher complexity load than the CAS scheme.   Imposing a complexity constraint causes the average complexity for successful decoding to increase, due to the occurrence of computational outages, which results in wasted computations and a corresponding increase in average complexity.  However, while this increase in complexity for the MRS is significant, especially at high SNR, the increase is barely noticeable for the CAS scheme. While there is still a chance of computational outage with the CAS scheme, the probability is too low to affect the average computational complexity.

\section{Outage in a Network}\label{sec:outage.network}
\begin{figure}[t]
\centering
\includegraphics[width=8.25cm]{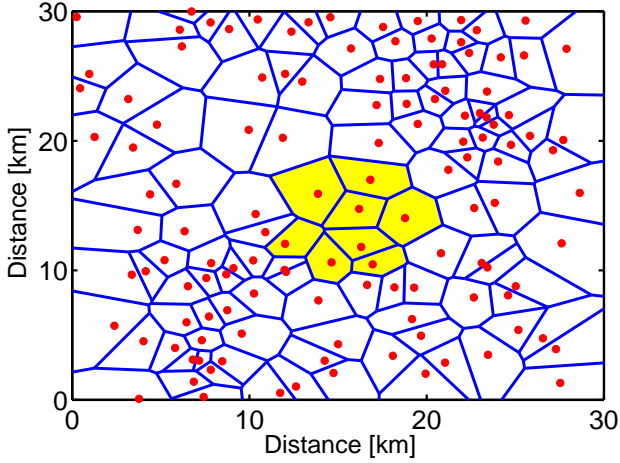}
\vspace{-0.25cm}
\caption{Base station locations. The cloud group consists of the eight highlighted cells in the center of the diagram. }
 \label{Figure:CellularNetwork}
   \vspace{-0.5cm}
\end{figure}

Now consider a network with $N_\mathsf{total}$ base stations or RAPs, where $N_\mathsf{total}$ could be arbitrarily large.  Unlike the single-cell scenario, there is co-channel interference on the uplink due to active mobiles in adjacent cells.  For illustrative purposes, we consider the network with $N_\mathsf{total}=129$ shown in Fig. \ref{Figure:CellularNetwork}, which is a segment of an actual deployment by a major provider in the UK at $1800$ MHz.  We consider the processing of the $N_\mathsf{cloud}=8$ cells highlighted in yellow.  Two options are considered, \emph{\ac{CP}}, where the signals from all $N_\mathsf{cloud}$ cells are processed in the cloud, and \emph{\ac{LP}}, where the signals at each RAP are locally processed.

We assume that the UEs are distributed according a \ac{PPP} with intensity $\lambda$ users per $\mathsf{km}^2$. The analysis of the cellular uplink with randomly distributed users has been recently considered in \cite{novlan:2013,torrieri:2013}.  While \cite{novlan:2013} assumes randomly placed base stations, the analysis in \cite{torrieri:2013} allows for arbitrarily placed base stations, such as those shown in Fig. \ref{Figure:CellularNetwork}.

RAP $Y_i$ serves a cell with area $\mathcal A_i$.  The number of UEs in $Y_i$'s cell is a Poisson variable with mean $\lambda \mathcal A_i$.  While the SC-FDMA uplink channel is partitioned into a plurality of \acp{RB} that could be allocated to different users, we assume that all of the available resources of a particular subframe are allocated to a single UE.  Computationally, this is a worst-case scenario as there is no computational diversity within a given cell.   Continuing the example from the previous section, we assume a 10 MHz system with 45 \acp{RB} allocated to each uplink user.  When there is more than one user in $Y_i$'s cell, then one of the users is selected at random to transmit on the uplink.   The likelihood of a transmitting user in $Y_i$'s cell is the complement of the void probability.  Thus, there will be a UE transmitting to $Y_i$ with probablility $1-\exp( - \lambda \mathcal A_i )$, and when there is a transmission, the location of the user is uniform within the cell.  The location of the UE transmitting to $Y_i$ is denoted by $X_i$.

The path loss from a mobile $X$ to a base station $Y$ is $|Y-X|^{-\alpha}$, where $\alpha$ is the \emph{path-loss exponent}.   $X_i$ transmits its uplink signal with power $P_i$.  The value of $P_i$ is selected according to the fractional power-control policy:
\begin{eqnarray}
   P_i
   & = &
   P_0 |Y_i - X_i|^{s \alpha}
\end{eqnarray}
where $P_0$ is a reference power (typically taken to be the power received at unit distance from the transmitter) and $s, 0 \leq s \leq 1,$ is the \emph{compensation factor} for fractional power control.  We assume $s=0.1$, which is the value reported in  \cite{coupechoux:2011} that maximizes the sum throughput. Higher values of $s$ would improve fairness at the cost of throughput.

The fading power gain from $X_i$ to $Y_j$ is $g_{i,j}$, which is normalized to have unit mean.  We again assume that $g_{i,j}$ is exponential (Rayleigh fading) and that the fading power gains remain fixed for the subframe, but vary from one subframe to the next (block fading). The SINR at $Y_j$ is
\begin{eqnarray}
  \gamma_j
  & = &
  \frac{ P_j g_{j,j} |Y_j-X_j|^{-\alpha} }{ W + \displaystyle \sum_{i \neq j} P_i g_{i,j} |Y_j-X_i|^{-\alpha}} \nonumber \\
  & = &
  \frac{  g_{j,j} |Y_j-X_j|^{\alpha(s-1)} }{ \Gamma^{-1} + \displaystyle \sum_{i \neq j} g_{i,j} |Y_j-X_i|^{-\alpha} |Y_i-X_i|^{s \alpha } } \label{Eqn:SINR}
\end{eqnarray}
where $W$ is the noise power, $\Gamma = P_0/W$ is the SNR at unit distance, and the summation is over all the co-channel interferers.

As in the single-cell case described in Section \ref{sec:outage.single.cell}, the throughput is determined with the assistance of a Monte Carlo simulation.  During each trial, which corresponds to a subframe, a mobile is placed at random in the $i^{th}$ cell with probability $1-\exp( - \lambda \mathcal A_i )$.  Once the mobiles are placed, the fading coefficients $g_{i,j}$ are drawn from an exponential distribution and the SINR at each RAP in the cloud group is computed according to (\ref{Eqn:SINR}).  The TBs that are in a channel outage are identified and the computational effort $C(\gamma_i)$ for each TB is determined as before.

Both \ac{LP} and \ac{CP} are considered.  In the \ac{LP} case, if $C(\gamma_i) > C_\mathsf{max}$ for a given TB, then it is considered to be in an outage.  For the \ac{CP} case, if (\ref{Eqn:Complexity}) is violated, then an outage occurs in at least one of the uplink TBs.  Scheduling in the virtual BS pool is assumed to process the signals with low SINR first.  This allows those TBs that are sent at a low MCS, and hence have a lower complexity requirement, to be first decoded, and the TBs that fail are the ones that were sent at the highest MCSs.

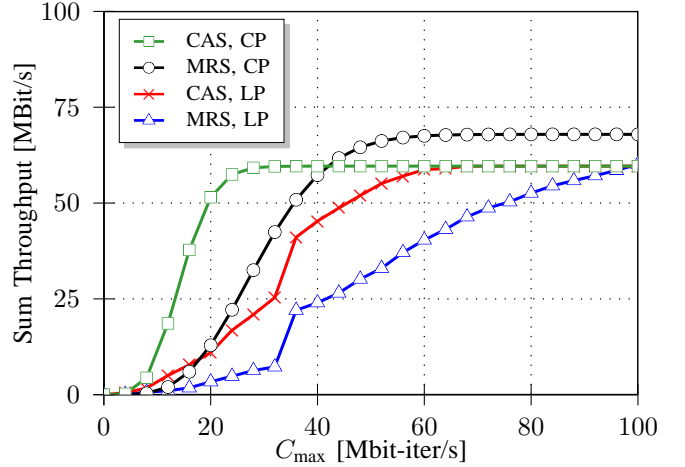
\begin{figure}[t]
\centering
\begingroup
\unitlength=1mm
\psset{xunit=0.71000mm, yunit=0.51000mm, linewidth=0.1mm}
\psset{arrowsize=2pt 3, arrowlength=1.4, arrowinset=.4}\psset{axesstyle=frame}
\begin{pspicture}(-19.71831, -27.45098)(100.00000, 100.00000)
\rput(-2.81690, -9.80392){%
\psaxes[subticks=0, labels=all, xsubticks=1, ysubticks=1, Ox=0, Oy=0, Dx=20, Dy=25]{-}(0.00000, 0.00000)(0.00000, 0.00000)(100.00000, 100.00000)%
\multips(20.00000, 0.00000)(20.00000, 0.0){4}{\psline[linecolor=black, linestyle=dotted, linewidth=0.2mm](0, 0)(0, 100.00000)}
\multips(0.00000, 25.00000)(0, 25.00000){3}{\psline[linecolor=black, linestyle=dotted, linewidth=0.2mm](0, 0)(100.00000, 0)}
\rput[b](50.00000, -17.64706){$C_\text{max}$ [Mbit-iter/s]}
\rput[t]{90}(-16.90141, 50.00000){Sum Throughput [MBit/s]}
\psclip{\psframe(0.00000, 0.00000)(100.00000, 100.00000)}
\psline[linecolor=blue, plotstyle=curve, linewidth=0.4mm, showpoints=true, linestyle=solid, linecolor=blue, dotstyle=triangle, dotscale=1.2 1.2, linewidth=0.4mm](0.00000, 0.00000)(4.00000, 0.10891)(8.00000, 0.65911)(12.00000, 1.02963)(16.00000, 1.79254)(20.00000, 3.38206)(24.00000, 4.81946)(28.00000, 6.31255)(32.00000, 7.26304)(36.00000, 22.04293)(40.00000, 24.00097)(44.00000, 26.53287)(48.00000, 30.16073)(52.00000, 32.99011)(56.00000, 37.10451)(60.00000, 40.37004)(64.00000, 43.16096)(68.00000, 46.41475)(72.00000, 48.73214)(76.00000, 50.33574)(80.00000, 52.60146)(84.00000, 54.50488)(88.00000, 55.86481)(92.00000, 57.14544)(96.00000, 58.52967)(100.00000, 60.11754)
\psline[linecolor=red, plotstyle=curve, linewidth=0.4mm, showpoints=true, linestyle=solid, linecolor=red, dotstyle=x, dotscale=1.2 1.2, linewidth=0.4mm](0.00000, 0.00000)(4.00000, 0.52994)(8.00000, 1.74382)(12.00000, 5.02025)(16.00000, 7.79400)(20.00000, 10.98048)(24.00000, 16.80991)(28.00000, 20.90328)(32.00000, 25.36450)(36.00000, 41.08766)(40.00000, 45.29039)(44.00000, 48.71078)(48.00000, 51.95452)(52.00000, 55.11779)(56.00000, 57.03462)(60.00000, 58.71326)(64.00000, 59.00239)(68.00000, 59.60388)(72.00000, 59.63311)(76.00000, 59.63313)(80.00000, 59.63409)(84.00000, 59.63413)(88.00000, 59.63413)(92.00000, 59.63413)(96.00000, 59.63413)(100.00000, 59.63413)
\psline[linecolor=black, plotstyle=curve, linewidth=0.4mm, showpoints=true, linestyle=solid, linecolor=black, dotstyle=o, dotscale=1.2 1.2, linewidth=0.4mm](0.00000, 0.00000)(4.00000, 0.01104)(8.00000, 0.35988)(12.00000, 1.99261)(16.00000, 5.98938)(20.00000, 12.89524)(24.00000, 22.15285)(28.00000, 32.49278)(32.00000, 42.42991)(36.00000, 50.86382)(40.00000, 57.29076)(44.00000, 61.73262)(48.00000, 64.54822)(52.00000, 66.19466)(56.00000, 67.08526)(60.00000, 67.53834)(64.00000, 67.75354)(68.00000, 67.85090)(72.00000, 67.89271)(76.00000, 67.90973)(80.00000, 67.91670)(84.00000, 67.91931)(88.00000, 67.92030)(92.00000, 67.92060)(96.00000, 67.92071)(100.00000, 67.92076)
\psline[linecolor=darkgreen, plotstyle=curve, linewidth=0.4mm, showpoints=true, linestyle=solid, linecolor=darkgreen, dotstyle=square, dotscale=1.2 1.2, linewidth=0.4mm](0.00000, 0.00000)(4.00000, 0.23313)(8.00000, 4.41350)(12.00000, 18.60080)(16.00000, 37.76939)(20.00000, 51.50688)(24.00000, 57.46341)(28.00000, 59.19728)(32.00000, 59.55747)(36.00000, 59.62009)(40.00000, 59.63196)(44.00000, 59.63394)(48.00000, 59.63413)(52.00000, 59.63413)(56.00000, 59.63413)(60.00000, 59.63413)(64.00000, 59.63413)(68.00000, 59.63413)(72.00000, 59.63413)(76.00000, 59.63413)(80.00000, 59.63413)(84.00000, 59.63413)(88.00000, 59.63413)(92.00000, 59.63413)(96.00000, 59.63413)(100.00000, 59.63413)
\endpsclip
\psframe[linecolor=black, fillstyle=solid, fillcolor=white, shadowcolor=lightgray, shadowsize=1mm, shadow=true](2.81690, 65.68627)(33.80282, 98.03922)
\rput[l](15.49296, 92.15686){\footnotesize{CAS, CP}}
\psline[linecolor=darkgreen, linestyle=solid, linewidth=0.3mm](5.63380, 92.15686)(11.26761, 92.15686)
\psline[linecolor=darkgreen, linestyle=solid, linewidth=0.3mm](5.63380, 92.15686)(11.26761, 92.15686)
\psdots[linecolor=darkgreen, linestyle=solid, linewidth=0.3mm, dotstyle=square, dotscale=1.2 1.2, linecolor=darkgreen](8.45070, 92.15686)
\rput[l](15.49296, 85.29412){\footnotesize{MRS, CP}}
\psline[linecolor=black, linestyle=solid, linewidth=0.3mm](5.63380, 85.29412)(11.26761, 85.29412)
\psline[linecolor=black, linestyle=solid, linewidth=0.3mm](5.63380, 85.29412)(11.26761, 85.29412)
\psdots[linecolor=black, linestyle=solid, linewidth=0.3mm, dotstyle=o, dotscale=1.2 1.2, linecolor=black](8.45070, 85.29412)
\rput[l](15.49296, 78.43137){\footnotesize{CAS, LP}}
\psline[linecolor=red, linestyle=solid, linewidth=0.3mm](5.63380, 78.43137)(11.26761, 78.43137)
\psline[linecolor=red, linestyle=solid, linewidth=0.3mm](5.63380, 78.43137)(11.26761, 78.43137)
\psdots[linecolor=red, linestyle=solid, linewidth=0.3mm, dotstyle=x, dotscale=1.2 1.2, linecolor=red](8.45070, 78.43137)
\rput[l](15.49296, 71.56863){\footnotesize{MRS, LP}}
\psline[linecolor=blue, linestyle=solid, linewidth=0.3mm](5.63380, 71.56863)(11.26761, 71.56863)
\psline[linecolor=blue, linestyle=solid, linewidth=0.3mm](5.63380, 71.56863)(11.26761, 71.56863)
\psdots[linecolor=blue, linestyle=solid, linewidth=0.3mm, dotstyle=triangle, dotscale=1.2 1.2, linecolor=blue](8.45070, 71.56863)
}\end{pspicture}
\endgroup
 
\caption{Sum throughput as function of the per-RAP complexity constraint when $N_\mathsf{cloud}=8$, with local processing (LP) and with cloud processing (CP).  Two MCS-selection schemes are considered: computationally aware selection (CAS) and max-rate selection (MRS). }
 \label{Figure:SumThroughputLorSprocessing}
\end{figure}


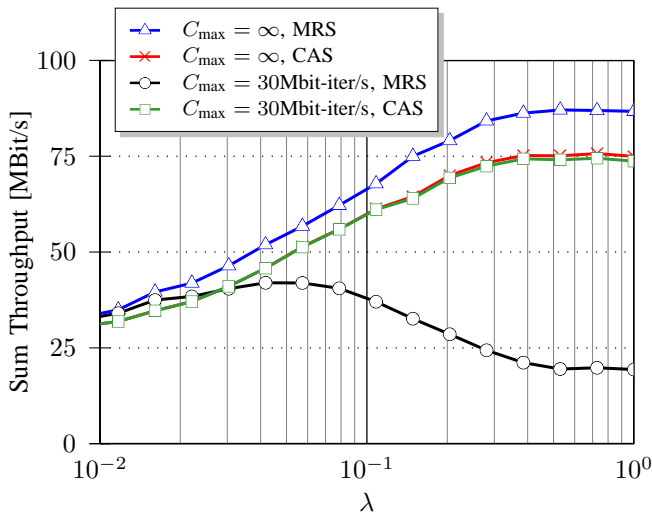
\begin{figure}[t]
\centering
\begingroup
\unitlength=1mm
\psset{xunit=35.50000mm, yunit=0.51000mm, linewidth=0.1mm}
\psset{arrowsize=2pt 3, arrowlength=1.4, arrowinset=.4}\psset{axesstyle=frame}
\begin{pspicture}(-2.39437, -27.45098)(0.00000, 100.00000)
\rput(-0.05634, -9.80392){%
\psaxes[subticks=10, xlogBase=10, logLines=x, labels=all, xsubticks=10, ysubticks=1, Ox=-2, Oy=0, Dx=1, Dy=25]{-}(-2.00000, 0.00000)(-2.00000, 0.00000)(0.00000, 100.00000)%
\multips(-2.00000, 25.00000)(0, 25.00000){3}{\psline[linecolor=black, linestyle=dotted, linewidth=0.2mm](0, 0)(2.00000, 0)}
\rput[b](-1.00000, -17.64706){$\lambda$}
\rput[t]{90}(-2.33803, 50.00000){Sum Throughput [MBit/s]}
\psclip{\psframe(-2.00000, 0.00000)(0.00000, 100.00000)}
\psline[linecolor=blue, plotstyle=curve, linewidth=0.4mm, showpoints=true, linestyle=solid, linecolor=blue, dotstyle=triangle, dotscale=1.2 1.2, linewidth=0.4mm](-4.00000, 26.42134)(-3.86207, 26.91080)(-3.72414, 26.81903)(-3.58621, 26.61474)(-3.44828, 26.93704)(-3.31034, 26.79957)(-3.17241, 26.81900)(-3.03448, 27.29994)(-2.89655, 27.78815)(-2.75862, 27.98687)(-2.62069, 28.93468)(-2.48276, 29.63109)(-2.34483, 30.37130)(-2.20690, 31.29997)(-2.06897, 32.98533)(-1.93103, 34.93006)(-1.79310, 39.58793)(-1.65517, 41.89916)(-1.51724, 46.42813)(-1.37931, 51.92708)(-1.24138, 56.74446)(-1.10345, 62.19275)(-0.96552, 67.82817)(-0.82759, 74.99815)(-0.68966, 79.08128)(-0.55172, 84.22718)(-0.41379, 86.24720)(-0.27586, 87.06993)(-0.13793, 86.93051)(0.00000, 86.70923)
\psline[linecolor=red, plotstyle=curve, linewidth=0.4mm, showpoints=true, linestyle=solid, linecolor=red, dotstyle=x, dotscale=1.2 1.2, linewidth=0.4mm](-4.00000, 24.93493)(-3.86207, 24.83848)(-3.72414, 24.75086)(-3.58621, 25.06201)(-3.44828, 25.39074)(-3.31034, 24.97390)(-3.17241, 25.08062)(-3.03448, 25.37223)(-2.89655, 25.59826)(-2.75862, 26.31370)(-2.62069, 26.08332)(-2.48276, 27.14847)(-2.34483, 28.05397)(-2.20690, 28.44732)(-2.06897, 30.66434)(-1.93103, 31.88932)(-1.79310, 34.67971)(-1.65517, 37.08791)(-1.51724, 41.03189)(-1.37931, 45.79641)(-1.24138, 51.39591)(-1.10345, 55.97316)(-0.96552, 61.23841)(-0.82759, 64.48080)(-0.68966, 70.00256)(-0.55172, 73.30197)(-0.41379, 75.18094)(-0.27586, 75.10099)(-0.13793, 75.69142)(0.00000, 74.95811)
\psline[linecolor=black, plotstyle=curve, linewidth=0.4mm, showpoints=true, linestyle=solid, linecolor=black, dotstyle=o, dotscale=1.2 1.2, linewidth=0.4mm](-4.00000, 26.41997)(-3.86207, 26.90737)(-3.72414, 26.81533)(-3.58621, 26.60791)(-3.44828, 26.93284)(-3.31034, 26.79113)(-3.17241, 26.80936)(-3.03448, 27.27864)(-2.89655, 27.73657)(-2.75862, 27.93464)(-2.62069, 28.83403)(-2.48276, 29.49819)(-2.34483, 30.09860)(-2.20690, 30.95395)(-2.06897, 32.33404)(-1.93103, 34.03888)(-1.79310, 37.48293)(-1.65517, 38.42081)(-1.51724, 40.43648)(-1.37931, 41.97759)(-1.24138, 41.94198)(-1.10345, 40.53513)(-0.96552, 37.04692)(-0.82759, 32.60870)(-0.68966, 28.55495)(-0.55172, 24.37442)(-0.41379, 21.15340)(-0.27586, 19.48432)(-0.13793, 19.80667)(0.00000, 19.36045)
\psline[linecolor=darkgreen, plotstyle=curve, linewidth=0.4mm, showpoints=true, linestyle=solid, linecolor=darkgreen, dotstyle=square, dotscale=1.2 1.2, linewidth=0.4mm](-4.00000, 24.93493)(-3.86207, 24.83848)(-3.72414, 24.75086)(-3.58621, 25.06201)(-3.44828, 25.39074)(-3.31034, 24.97390)(-3.17241, 25.08062)(-3.03448, 25.37223)(-2.89655, 25.59826)(-2.75862, 26.31370)(-2.62069, 26.08332)(-2.48276, 27.14847)(-2.34483, 28.05397)(-2.20690, 28.44732)(-2.06897, 30.66434)(-1.93103, 31.88932)(-1.79310, 34.67955)(-1.65517, 37.08783)(-1.51724, 41.03118)(-1.37931, 45.79215)(-1.24138, 51.29092)(-1.10345, 55.93696)(-0.96552, 61.06807)(-0.82759, 63.98708)(-0.68966, 69.35694)(-0.55172, 72.40282)(-0.41379, 74.31623)(-0.27586, 74.06318)(-0.13793, 74.48018)(0.00000, 73.70713)
\endpsclip
\psframe[linecolor=black, fillstyle=solid, fillcolor=white, shadowcolor=lightgray, shadowsize=1mm, shadow=true](-1.94366, 81.37255)(-0.73239, 113.72549)
\rput[l](-1.69014, 107.84314){\footnotesize{$C_\text{max}= \infty$, MRS}}
\psline[linecolor=blue, linestyle=solid, linewidth=0.3mm](-1.88732, 107.84314)(-1.77465, 107.84314)
\psline[linecolor=blue, linestyle=solid, linewidth=0.3mm](-1.88732, 107.84314)(-1.77465, 107.84314)
\psdots[linecolor=blue, linestyle=solid, linewidth=0.3mm, dotstyle=triangle, dotscale=1.2 1.2, linecolor=blue](-1.83099, 107.84314)
\rput[l](-1.69014, 100.98039){\footnotesize{$C_\text{max}= \infty$, CAS}}
\psline[linecolor=red, linestyle=solid, linewidth=0.3mm](-1.88732, 100.98039)(-1.77465, 100.98039)
\psline[linecolor=red, linestyle=solid, linewidth=0.3mm](-1.88732, 100.98039)(-1.77465, 100.98039)
\psdots[linecolor=red, linestyle=solid, linewidth=0.3mm, dotstyle=x, dotscale=1.2 1.2, linecolor=red](-1.83099, 100.98039)
\rput[l](-1.69014, 94.11765){\footnotesize{$C_\text{max}= 30\text{Mbit-iter/s}$, MRS}}
\psline[linecolor=black, linestyle=solid, linewidth=0.3mm](-1.88732, 94.11765)(-1.77465, 94.11765)
\psline[linecolor=black, linestyle=solid, linewidth=0.3mm](-1.88732, 94.11765)(-1.77465, 94.11765)
\psdots[linecolor=black, linestyle=solid, linewidth=0.3mm, dotstyle=o, dotscale=1.2 1.2, linecolor=black](-1.83099, 94.11765)
\rput[l](-1.69014, 87.25490){\footnotesize{$C_\text{max}= 30\text{Mbit-iter/s}$, CAS}}
\psline[linecolor=darkgreen, linestyle=solid, linewidth=0.3mm](-1.88732, 87.25490)(-1.77465, 87.25490)
\psline[linecolor=darkgreen, linestyle=solid, linewidth=0.3mm](-1.88732, 87.25490)(-1.77465, 87.25490)
\psdots[linecolor=darkgreen, linestyle=solid, linewidth=0.3mm, dotstyle=square, dotscale=1.2 1.2, linecolor=darkgreen](-1.83099, 87.25490)
}\end{pspicture}
\endgroup
 
\caption{Sum throughput with cloud processing as function of the density of UEs when $N_\mathsf{cloud}=8$. Two MCS-selection schemes are considered: computationally aware selection (CAS) and max-rate selection (MRS). }
 \label{Figure:SumThroughputLambda}
\end{figure}

Fig. \ref{Figure:SumThroughputLorSprocessing} shows the sum throughput, which is the sum of the effective throughputs in each cell of the cloud group, as a function of the per-RAP complexity constraint $C_\mathsf{max}$ with the two MCS-selection policies.  The plot was generated assuming a UE density of $\lambda = 0.1$, path-loss exponent $\alpha = 3.7$, and average SNR $\Gamma = 20$ dB.   As can be seen, \ac{CP} is much more computationally efficient than \ac{LP}. For the same complexity constraint, the throughput of \ac{CP} is significantly higher than it is with \ac{LP}.  The computational diversity advantage of \ac{CP} is evident from its steeper curve.   Alternatively, the complexity constraint required to achieve a desired throughput is much lower for \ac{CP} than it is for \ac{LP}.  These behaviors confirm the computational diversity benefits of cloud processing.  Furthermore, there is a benefit to using the more conservative CAS policy  for the \ac{LP} system.  For the \ac{CP} system, CAS is beneficial as long as $C_\mathsf{max} \leq 40$ Mbit-iter/s; for systems with less of a complexity constraint MRS is better.



Fig. \ref{Figure:SumThroughputLambda} shows the effect of mobile density $\lambda$ on the sum throughput.  Again, $\alpha = 3.7$, and the average SNR $\Gamma = 20$ dB. The figure shows results for \ac{CP}, both with no complexity constraint ($C_\mathsf{max}=\infty$) and with a constraint of $C_\mathsf{max} = 30$ Mbit-iter/s.  The two MCS selection policies are again considered.  When there is no complexity constraint, the sum throughput rises with $\lambda$.  In this case, the throughput of MRS is higher than CAS.  However, when there is a complexity constraint, throughput falls with higher values $\lambda$ when MRS is used. This is due to a high prevalence of computational outages in heavily loaded systems.  However, if the more conservative CAS policy is used, performance is approximately the same as if there were no complexity constraint.  This confirms the effectiveness of using a more conservative MCS-selection policy in dense networks.


%

\balance
\section{Conclusions}\label{sec:conclusions}
Computational limitations in mobile wireless networks have a significant impact on the network performance in the moderate to high \ac{SNR} regime. Such computational limitations are of particular interest to small-cell networks, where \acp{RAP} are computationally limited due to economic constraints.  The computational constraints are also of fundamental importance to the design of centralized RAN architectures.    With the framework introduced in this paper, a direct link between the employed computing technology and the communication technology can be drawn, which opens the opportunity to exploit new degrees of freedom within a mobile network.

The newly introduced concept of \emph{computational outage} helps to quantify the complexity-throughput tradeoff in centralized RAN platforms.  Such a quantification allows the computational requirements for centralized RAN platforms to be specified in a meaningful way.  The analysis furthermore provides insight into the \emph{computational diversity} that can be achieved by cloud processing, and the corresponding increases in effective throughput.

This paper discussed a single-cell and a multi-cell scenario where inter-cell interference has a considerable impact on the network performance.  It was found that in computationally constrained mobile networks, there is a benefit to intentionally selecting a lower MCS value, as its lower computational demand may actually increase the effective throughput due to a reduction in computational outage.  The benefits of using cloud processing and using a conservative MCS selection policy become more pronounced as the user density increases.

The analysis in this paper has assumed block-fading channels.  In practice, user mobility and time-variant channels will have
  a significant impact on the performance of a centralized RAN.  The study of such systems is a key future direction for this research.  It may be that the solution is to use an aggressive MCS-selection scheme for some users, and reserve the conservative MCS-selection scheme for only a small number of user so as to not overload the system.
 Such a strategy could be interpreted as a kind of ``computational water-filling'', and could allow for reducing the ratio of peak-to-average computational effort in a
centralized RAN, thereby increasing the utilization of the network infrastructure.

  \bibliographystyle{IEEEtran}
  \bibliography{IEEEfull,references}
\end{document}